\documentstyle[12pt,a4,epsf]{article}
\newdimen\unit
\def\point#1 #2 #3{\vbox to0pt{\kern-#2\unit
  \hbox{\kern#1\unit$#3$}\vss}
 \nointerlineskip}
\def\b2rho{\bar B^0\to\rho^+ l^- \bar\nu_l}
\def\btopi{\bar B^0\to\pi^+ l^- \bar\nu_l}
\def\btokstargamma{\bar B\to K^* \gamma}
\def\vub{|V_{ub}|}
\def\qsqmax{q^2_{\rm max}}
\def\w{\omega}
\def\gev{\,{\rm Ge\kern-0.1em V}}
\def\mev{\,{\rm Me\kern-0.1em V}}
\makeatletter
\def\big#1{{\hbox{$\left#1\vbox to1.012\ht\strutbox{}\right.\n@space$}}}
\def\Big#1{{\hbox{$\left#1\vbox to1.369\ht\strutbox{}\right.\n@space$}}}
\def\bigg#1{{\hbox{$\left#1\vbox to1.726\ht\strutbox{}\right.\n@space$}}}
\def\Bigg#1{{\hbox{$\left#1\vbox
to2.083\ht\strutbox{}\right.\n@space$}}}
\long\def\@makecaption#1#2{
 \vskip 15pt 
 \small
 \setbox\@tempboxa\hbox{{\bf #1} \ #2}
 \ifdim \wd\@tempboxa >\hsize \unhbox\@tempboxa\par \else \hbox
to\hsize{\hfil\box\@tempboxa\hfil}
 \fi}
\makeatother
\def\tstrut{\vrule height2.5ex depth0pt width0pt}

\begin{document}
\begin{titlepage}
\begin{flushright}
Southampton Preprint SHEP 95--18\\
Edinburgh Preprint 95/551\\
Marseille Preprint CPT--95/PE.3218\\
hep-ph/9506398
\end{flushright}

\bigskip

\begin{center}
{\Huge Lattice Study of the Decay $\bar B^0\to\rho^+ l^- \bar\nu_l$:\\
Model-Independent Determination of $|V_{ub}|$}\\[5em]
{\large\it UKQCD Collaboration}\\[1.5ex]

{\bf J~M~Flynn, J~Nieves}\\
Department of Physics, University of Southampton, Southampton
SO17~1BJ, UK\\[1ex]
{\bf K~C~Bowler, N~M~Hazel, D~S~Henty, H~Hoeber\footnote{Present
address: HLRZ, 52425 J\"ulich and DESY, Hamburg, Germany},
R~D~Kenway and D~G~Richards}\\
Department of Physics \& Astronomy, The University of
Edinburgh, Edinburgh EH9~3JZ, Scotland\\[1ex]
{\bf B~J~Gough}\\
Theoretical Physics MS106, Fermilab, Batavia IL 60510, USA\\[1ex]
{\bf H~P~Shanahan}\\
Department of Physics \& Astronomy, University of Glasgow, Glasgow
G12~8QQ, Scotland\\[1ex]
{\bf L~P~Lellouch}\\
Centre de Physique Theorique, CNRS Luminy,
F--13288 Marseille Cedex 9, France\footnote{Unit\'e Propre de
Recherche 7061}
\end{center}

\bigskip

\begin{abstract}
We present results of a lattice computation of the vector and
axial-vector current matrix elements relevant for the semileptonic
decay $\b2rho$. The computations are performed in the quenched
approximation of lattice QCD on a $24^3\times 48$ lattice at $\beta =
6.2$, using an ${\cal O}(a)$ improved fermionic action.  Our principal
result is for the differential decay rate, $d\Gamma/dq^2$, for the
decay $\b2rho$ in a region beyond the charm threshold, allowing a
model-independent extraction of $\vub$ from experimental
measurements. Heavy quark symmetry relations between radiative and
semileptonic decays of $\bar B$ mesons into light vector mesons are
also discussed.
\end{abstract}

\end{titlepage}

\section{Introduction}

The experimental determination of the Cabibbo-Kobayashi-Maskawa
(CKM)~\cite{buras-harlander:ckm,alilondon:ichep94} matrix elements is
vital because they control the hadronic sector of the Standard Model,
determining CP violation and flavour mixing. The magnitudes and phases
of these matrix elements must be established to make the Standard
Model predictive, to determine the angles and area of the unitarity
triangle, and to look for any inconsistency between the Standard Model
and the experimental data which would point to new physics.

The CKM element $V_{ub}$ is one of the most poorly known, currently
uncertain to within a factor of two or three (at 90\% CL) in
magnitude~\cite{pdg94}. The determination of $\vub$ has traditionally
been made from inclusive $b\to u$ semileptonic
decays~\cite{cleo:inclusive-b-to-u}, looking at the lepton energy
spectrum beyond the endpoint for decays to charmed final states. These
determinations rely on models incorporating nonperturbative QCD
effects to relate the measured spectrum to theoretical
predictions. Models used include the quark
model~\cite{acm:partonmodel} and bound-state
models~\cite{gisw}--\cite{ks} together with attempts to combine
features from both~\cite{rdb}. This results in the inclusion of an
additional error reflecting the range of models
used~\cite{pdg:vub-discussion}.

In this paper we propose a model-independent method to determine
$\vub$ from the exclusive semileptonic charmless $B$ decay $\b2rho$,
which should be measured with improved accuracy in $B$ factories and
at $e^+e^-$ and hadron colliders in the near future. Currently, only
an upper bound exists~\cite{cleo:b-to-rho} for the total decay rate,
but new experimental results should be available
soon~\cite{cleo:gibbons}.

We will show how lattice QCD calculations, which incorporate
nonperturbative QCD effects in a systematic way, can be used to
extract $\vub$ from experimental measurements of $\b2rho$. Exclusive
$B$ decays have already proved useful for extracting the value of
$|V_{cb}|$ by studying the process $\bar B\to D^* l
\bar\nu_l$~\cite{neubert:ichep94}%
--\cite{argus:btodstar}.

We have also analysed relations between the exclusive processes
$\btokstargamma$ and $\b2rho$ which follow from heavy quark symmetry
(HQS)~\cite{isgurwise:hqet}. We discuss the utility of these relations
for determining $\vub$ from the experimental measurements of ${\rm
B}(\btokstargamma)$ and ${\rm B}(b\to s\gamma)$~\cite{bd}%
--\cite{cds}.

\section{Form Factors}

The matrix elements we will be considering are of the $V-A$ weak
current between $\bar B$ and $\rho$ mesons and of the magnetic moment
operator between $\bar B$ and $K^*$ mesons~\cite{gsw}. For $\b2rho$
the matrix element is,
\begin{equation}
\langle \rho(k,\eta)| \bar u \gamma_\mu(1-\gamma_5)b| \bar B(p) \rangle
  = \eta^{*\beta} T_{\mu \beta}, \label{eq:btorho1}
\end{equation}
with form factor decomposition,
\begin{eqnarray}
T_{\mu \beta} &=& \frac{2V(q^2)}{m_B+m_\rho}
\epsilon_{\mu \gamma \delta \beta}p^{\gamma}k^{\delta}
- i(m_B+m_\rho)A_1(q^2)g_{\mu \beta} \nonumber \\
& & \mbox{} + i \frac{A_2(q^2)}{m_B+m_\rho} (p+k)_{\mu}q_{\beta} -
i \frac{A(q^2)}{q^2}2m_\rho q_{\mu}(p+k)_{\beta},  \label{eq:btorho2}
\end{eqnarray}
where $q = p - k$ is the four-momentum transfer and $\eta$ is the
$\rho$ polarisation vector. The form factor $A$ can be written as
\begin{equation}
A(q^2) = A_0(q^2)-A_3(q^2),
\end{equation}
where,
\begin{equation}
A_3(q^2) = \frac{m_B+m_\rho}{2m_\rho}A_1(q^2)-
\frac{m_B-m_\rho}{2m_\rho}A_2(q^2),
\end{equation}
with $A_0(0)=A_3(0)$. In the limit of zero lepton masses, the term
proportional to $A$ in equation~(\ref{eq:btorho2}) does not contribute
to the total amplitude and hence to the decay rates. Pole dominance
models suggest that $V$, $A_i$ for $i=1,2,3$ and $A_0$ correspond to
$1^-$, $1^+$ and $0^-$ exchanges respectively in the
$t$-channel~\cite{bsw}.

The main contribution to the $\btokstargamma$ decay comes from the
matrix element
\begin{equation}
\label{eq:btokstargamma}
\langle K^*(k,\eta) | \overline{s} \sigma_{\mu\nu} q^\nu b_R
 | \bar B(p) \rangle
  =  \sum_{i=1}^3 C^i_\mu T_i(q^2),
\end{equation}
where $q=p-k$ as above, $\eta$ is now the $K^*$ polarisation vector and
\begin{eqnarray}
C^{1}_\mu & = &
  2 \epsilon_{\mu\nu\lambda\rho} \eta^\nu p^\lambda k^\rho, \\
C^{2}_\mu & = &
  \eta_\mu(m_B^2 - m_{K^*}^2) - \eta\cdot q (p+k)_\mu, \\
C^{3}_\mu & = & \eta\cdot q
  \left( q_\mu - \frac{q^2}{m_B^2-m_{K^*}^2} (p+k)_\mu \right).
\end{eqnarray}
For an on-shell photon with $q^2=0$, $T_3$ does not contribute to the
$\btokstargamma$ amplitude and $T_1$ and $T_2$ are related by,
\begin{equation}
T_1(q^2{=}0) = i T_2(q^2{=}0).
\end{equation}
Hence, for $\btokstargamma$, we need to determine $T_1$ and/or $T_2$
at the on-shell point.

Neglecting corrections suppressed by inverse powers of the heavy quark
mass $M$, the following relations hold when $q^2$ is close to the
maximum recoil value $\qsqmax = (M-
m_{\rho,K^*})^2$~\cite{isgurwise:hqet}
\begin{equation}
V\Theta/\sqrt M = {\rm const}, \qquad
A_1\Theta \sqrt M = {\rm const}, \qquad
A_2\Theta/\sqrt M = {\rm const},\label{eq:btorho-infmass}
\end{equation}
where $\Theta$ arises from the leading logarithmic corrections and is
chosen to be 1 at the $B$ mass~\cite{neubert:physrep},
\begin{equation}\label{eq:theta}
\Theta = \Theta(M/m_B) = \left( \frac{\alpha_s(M)}{\alpha_s(m_B)}
                         \right)^{\frac{2}{\beta_0}}.
\end{equation}
In the calculations reported below, we will use $\beta_0 = 11$ in the
quenched approximation and $\Lambda_{\rm QCD} = 200\mev$. The matrix
elements of the $V-A$ current and the magnetic moment operator between
$\bar B$ and identical\footnote{We assume isospin symmetry between the
light $u$ and $d$ quarks.} light final-state vector mesons of mass $m$
are related in the infinite heavy quark mass limit according to:
\begin{equation}
V(q^2) = 2T_1(q^2), \qquad A_1(q^2) = 2i T_2(q^2),\label{eq:VT1andAT2}
\end{equation}
for values of $q^2$ not too far from $\qsqmax$, or equivalently, for
$\w$ close to $1$, where
\begin{equation}
\w = v \cdot v' = {M^2 + m^2 - q^2\over 2 M m}.
\end{equation}
Here, $v$ and $v'$ are the four-velocities of the heavy-light
pseudoscalar meson and the light vector meson respectively.

The equations in~(\ref{eq:VT1andAT2}) relate the physical decay
processes $\b2rho$ and $\bar B^- \to \rho^- \gamma$. If light flavour
$SU(3)$ symmetry is respected then equation~(\ref{eq:VT1andAT2})
relates the processes $\b2rho$ and
$\btokstargamma$~\cite{isgurwise:hqet}. In the lattice calculations
reported below we will test equation~(\ref{eq:VT1andAT2}) directly for
different values of the heavy quark mass using identical light meson
states for both matrix elements.

In the literature, versions of equation~(\ref{eq:VT1andAT2}) appear
which rely on lowest order HQS but incorporate all $1/M$ corrections
from kinematics~\cite{isgurwise:hqet,bd,GMM:form-factors}\footnote{To
incorporate properly the dynamical $1/M$ corrections it is necessary
to consider operators of dimension four in the heavy quark expansion
in addition to those of dimension three occurring at leading order.}.
The modified relations become:
\begin{eqnarray}
2T_1(q^2) &=& {q^2 + M^2 - m^2 \over 2 M}{V(q^2)\over M+m} +
              {M+m\over 2M} A_1(q^2) \label{eq:VT1kinematic} \\[1ex]
2iT_2(q^2) &=& {\big[(M{+}m)^2 - q^2\big]
                 \big[(M{-}m)^2 - q^2\big]\over2M(M^2{-}m^2)}
               {V(q^2)\over M+m} \nonumber \\
           & & \mbox{} +
              {q^2 + M^2 - m^2 \over 2 M}{A_1(q^2)\over M - m}
               \label{eq:AT2kinematic}
\end{eqnarray}

\section{Lattice Details}\label{sec:latticedetails}

Lattice calculations with propagating quarks provide matrix elements
for heavy quarks around the charm mass over a range of $q^2$
straddling $q^2=0$. In extracting form factors from these matrix
elements, we can reach $\qsqmax$ for $A_1$ and $T_2$ only, because the
coefficients determining the contribution of the other form factors to
the matrix elements of equations~(\ref{eq:btorho1})
and~(\ref{eq:btokstargamma}) vanish at this kinematic point.  To
obtain results relevant for $B$ decays, we need to extrapolate in the
heavy quark mass $M$ to the $b$ quark mass. This is simple for fixed
$\w$, but, at the $B$ scale, produces a range of $q^2$ values near
$\qsqmax$ and far from $q^2 = 0$.

The results described below come from $60$ $SU(3)$ gauge
configurations generated by the UKQCD collaboration on a $24^3 \times
48$ lattice at $\beta = 6.2$ in the quenched approximation. The ${\cal
O}(a)$ improved Sheikholeslami--Wohlert (SW)~\cite{sw-action} action
was used for fermions, with ``rotated'' fermion fields appearing in
all operators used for correlation function
calculations~\cite{heatlie:clover-action}. The inverse lattice spacing
determined from the $\rho$ mass is $a^{-1} =
2.7(1)\gev$~\cite{ukqcd:strange-prd}. Other physical quantities will
lead to slightly different values for the lattice spacing ($a^{-1} =
2.7$--$3.0 \gev$~\cite{ukqcd:fb-how}). The scale uncertainty should be
reflected in the results for dimensionful quantities.

\begin{figure}
\hbox to\hsize{\hss\vbox{\offinterlineskip
\epsfxsize=0.6\hsize\epsffile{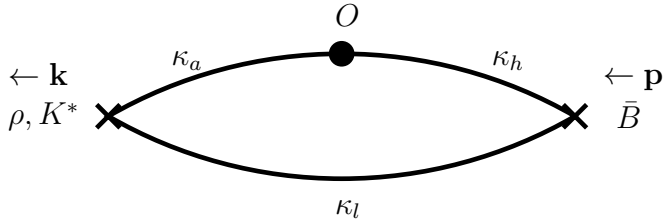}
\unit=0.6\hsize
\point -0.01 0.2 {\rho, K^*}
\point 0.92 0.2 {\bar B}
\point 0.49 0.35 O
\point 0.49 0.05 {\kappa_l}
\point 0.24 0.278 {\kappa_a}
\point 0.73 0.278 {\kappa_h}
\point 0.9 0.25 {\leftarrow \bf p}
\point -0.01 0.26 {\leftarrow \bf k}
}
\hss}
\caption[]{Labelling of quark hopping parameters for three-point
correlator calculation.}
\label{fig:threeptcorrelator}
\end{figure}

Three-point correlators of the heavy-to-light two-fermion operators
with a heavy pseudoscalar meson (the ``$\bar B$'' meson) and a light
vector meson were calculated, as illustrated in
figure~\ref{fig:threeptcorrelator}. Matrix elements were extracted
from these correlators by the method detailed
in~\cite{ukqcd:bsg2}--%
\cite{bks:dtok}. Four heavy-quark hopping parameters, $\kappa_h =
0.121, 0.125, 0.129, 0.133$, were used. For the propagator connecting
the current operator to the light meson operator, two kappa values,
$\kappa_a = 0.14144, 0.14226$, were available. The subscript $a$ is
for ``active'' to contrast this propagator with the ``spectator''
propagator joining the heavy-light meson to the light meson. These
$\kappa_a$ values straddle that for the strange quark,
$0.1419(1)$~\cite{ukqcd:strange-prd}. For $\kappa_h = 0.121, 0.129$,
we used three light spectator hopping parameters, $\kappa_l = 0.14144,
0.14226, 0.14262$, and for $\kappa_h = 0.125, 0.133$ we used $\kappa_l
= 0.14144$ only. The critical hopping parameter at this $\beta$ is
$\kappa_{\rm crit} = 0.14315(1)$~\cite{ukqcd:strange-prd}.

The lattice calculations were performed with the heavy meson spatial
momentum of magnitude $0$ or $1$, in lattice units of $\pi/12 a$. The
momentum injected at the operator insertion was varied to allow the
modulus of the light meson spatial momentum to take values up to
$\sqrt2$ in lattice units (although some of the momentum choices were
too noisy to be used in fits). We refer to each combination of heavy
and light meson three-momenta as a channel with the notation $|{\bf
p}|\to |{\bf k}|$ in lattice units (for example $0\to1$ or
$1\to1_\perp$ where the subscript $\perp$ indicates that $\bf p$ and
$\bf k$ are perpendicular).

The results below have been obtained using uncorrelated fits for the
extrapolations in the heavy quark mass. The extraction of the form
factors from the three-point correlation function data used correlated
fits~\cite{ukqcd:bsg2,ukqcd:dtok}. Statistical errors are 68\%
confidence limits obtained from $250$ bootstrap samples.

To make the best use of HQS, we pick momentum combinations which keep
$\w$ constant as the heavy-light meson mass varies. This allows us to
scale in $1/M$ from the charm to the bottom mass scale and gives the
form factors for the $\bar B$ decays as functions of $\w$, and
therefore as functions of $q^2$.  When the heavy-light meson is at
rest, $\w$ is independent of the heavy-light meson mass. As shown in
reference~\cite{ukqcd:hlff} there are some additional channels, where
the heavy-light meson is not at rest, but $\bf p$ and $\bf k$ are
perpendicular, for which $\w$ is very nearly constant as the heavy
meson mass varies, and which can be used for heavy quark mass
extrapolations. For those channels where $\w$ is not strictly constant
but which we nevertheless use in our study, the difference between the
average $\w$ value used in our analysis and that for each value of the
heavy quark mass is always less than 3\%~\cite{ukqcd:hlff}.

For relating lattice results to the continuum, we have used the
perturbative values for the renormalisation constants of the vector,
axial and magnetic moment operators~\cite{zva}:
\begin{equation}
Z_V = 0.83, \qquad Z_A = 0.97, \qquad Z_\sigma = 0.98.
\end{equation}

The SW improved action reduces the leading discretisation errors from
${\cal O} (a)$ in the Wilson fermion action to ${\cal O}(\alpha_s a)$,
but for quark masses $m_Q$ around that of the charm quark, $\alpha_s
m_Q a$ can be of order 10\%. An estimate of the lattice artefacts can
be obtained by comparing values of lattice renormalisation constants,
computed from different matrix elements, in a non-perturbative
way~\cite{npt:msv}.  Numerical calculations have confirmed that errors
of order 10\% are present at our value of $\beta$ in the matrix
elements of vector and axial vector
currents~\cite{ukqcd:btodstar,ukqcd:lpllat94}. We will therefore allow
for an extra 10\% systematic uncertainty in our results due to
possible discretisation errors.

\section{Semileptonic and Radiative Decays of Heavy-Light Mesons and
Heavy Quark Symmetry}

In this section we study the relations between the semileptonic and
radiative decay processes of heavy-light mesons to light vector
mesons. We will check the HQS relations of
equation~(\ref{eq:VT1andAT2}) and the size of $1/M$ corrections to
these relations for different values of the heavy-light meson
mass. For testing these relations we use our most accurate light-quark
data, $\kappa_l = \kappa_a = 0.14144$: this ensures that the light
degrees of freedom are the same for the semileptonic and radiative
processes and allows us to check directly, without corrections due to
$SU(3)$ symmetry breaking, the validity of HQS as $M$ increases.

Specifically, we compare the form factors $A_1$ and $V$ for the
pseudoscalar to vector semileptonic decay,
\begin{equation}
P_{hl}(\kappa_h;\kappa_l=0.14144) \to
  V_{ll}(\kappa_l=0.14144) l \bar\nu_l,
\end{equation}
with the form factors $T_1$ and $T_2$ for the pseudoscalar to vector
radiative decay,
\begin{equation}
P_{hl}(\kappa_h;\kappa_l=0.14144) \to
  V_{ll}(\kappa_l=0.14144) \gamma,
\end{equation}
for different values of the heavy quark mass, $\kappa_h = 0.121$,
$0.125$, $0.129$ and $0.133$, and for different values of $\w$ close
to $\w = 1$. The corresponding heavy-light pseudoscalar masses are in
the range $1.6\gev$ to $2.5\gev$. We have also extrapolated to the $B$
scale and to the infinite heavy quark mass limit. We have used five
momentum channels in our analysis\footnote{$A_1(\qsqmax)$ and
$T_2(\qsqmax)$ can be determined directly from the $0\to0$ data, but
$V(\qsqmax)$ and $T_1(\qsqmax)$ have to be obtained by an
extrapolation in $q^2$ from the measured data.}: $0\to0$, $0\to1$,
$0\to\sqrt2$, $1\to0$ and $1\to1_\perp$.

\begin{figure}
\vbox{%
\hbox to\hsize{\epsfxsize=0.48\hsize
\epsffile[37 37 510 510]{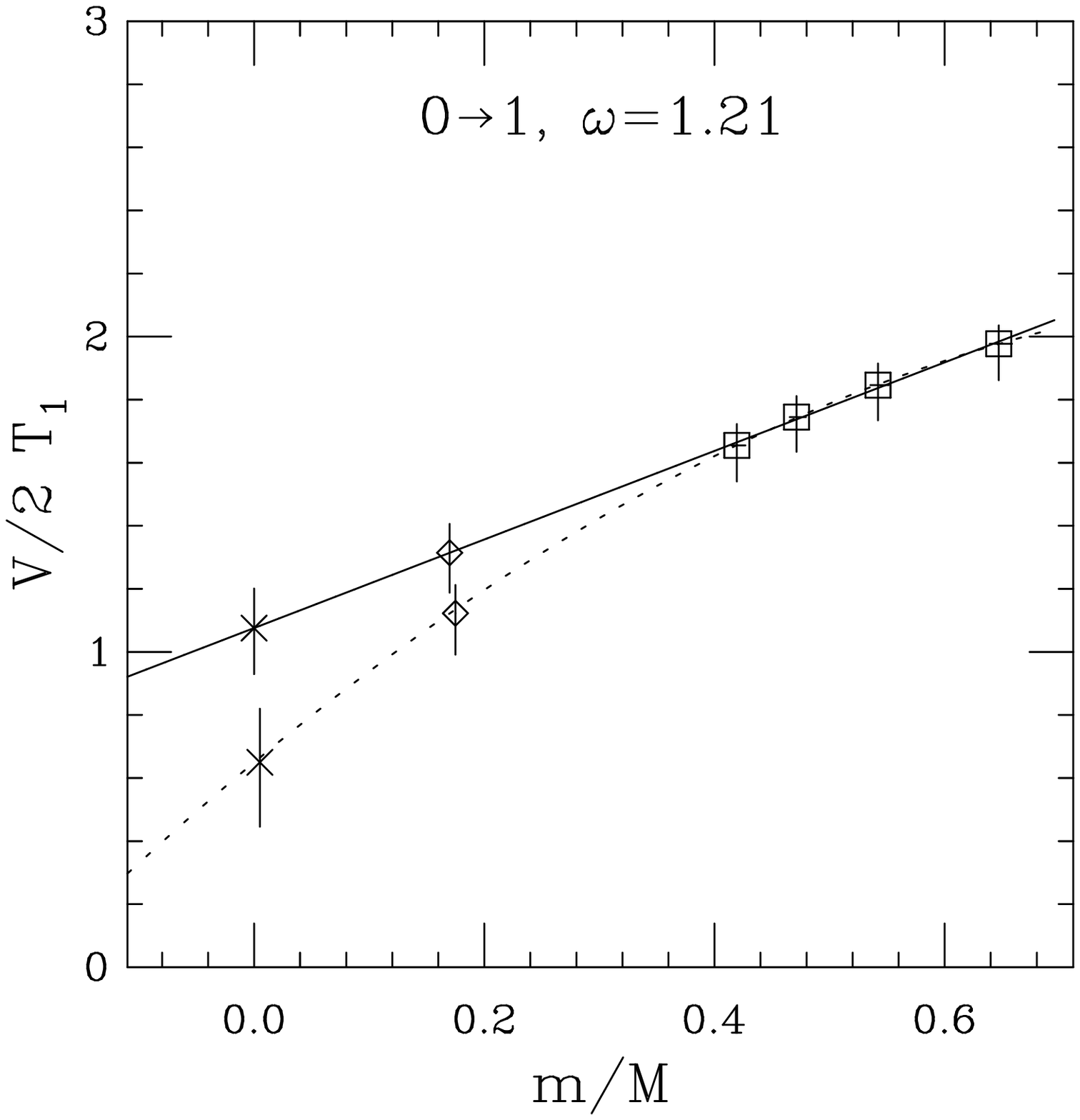}
\hfill\epsfxsize=0.48\hsize
\epsffile[37 37 510 510]{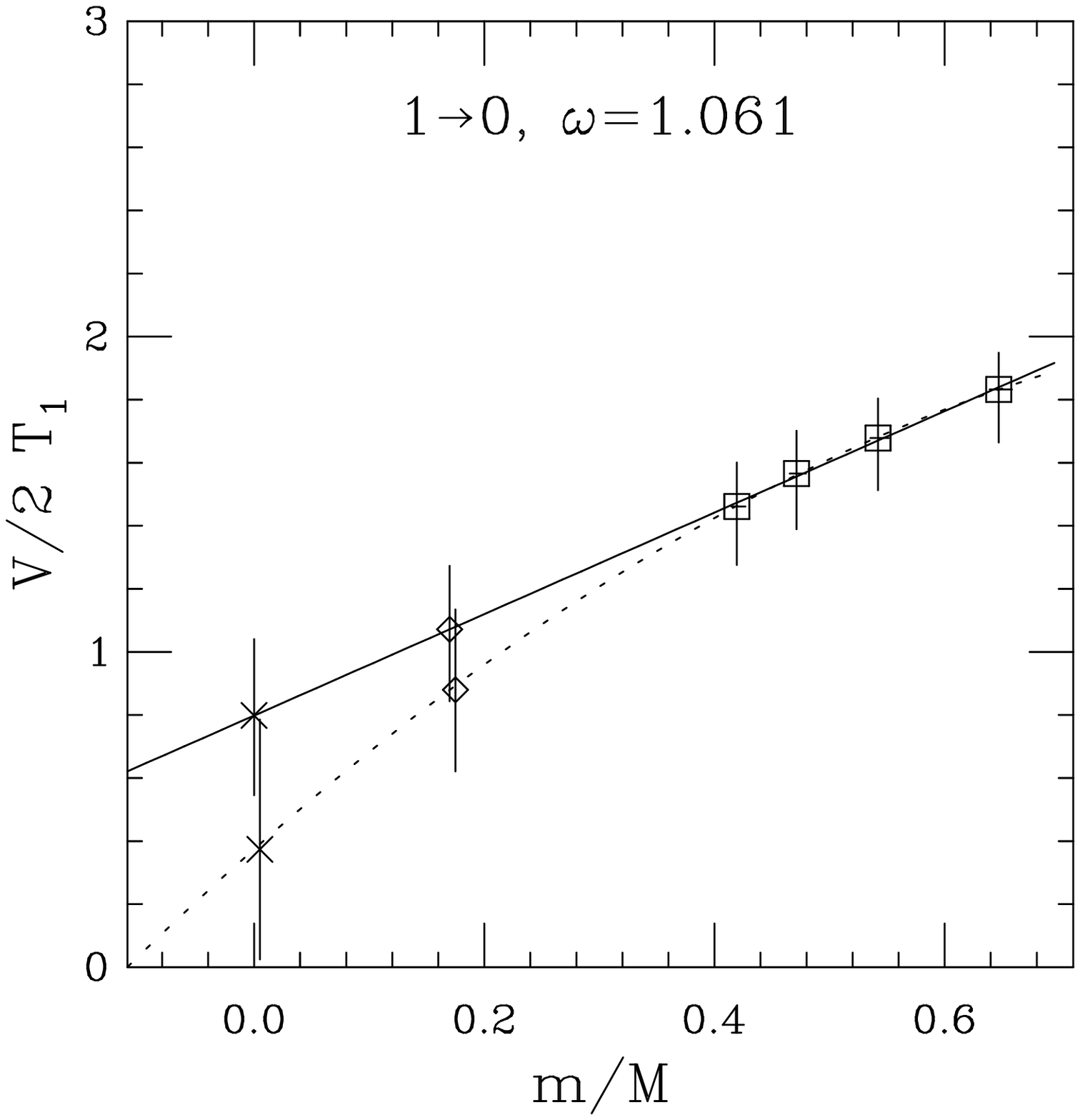}}
\kern1em
\hbox to\hsize{\epsfxsize=0.48\hsize
\epsffile[37 37 510 510]{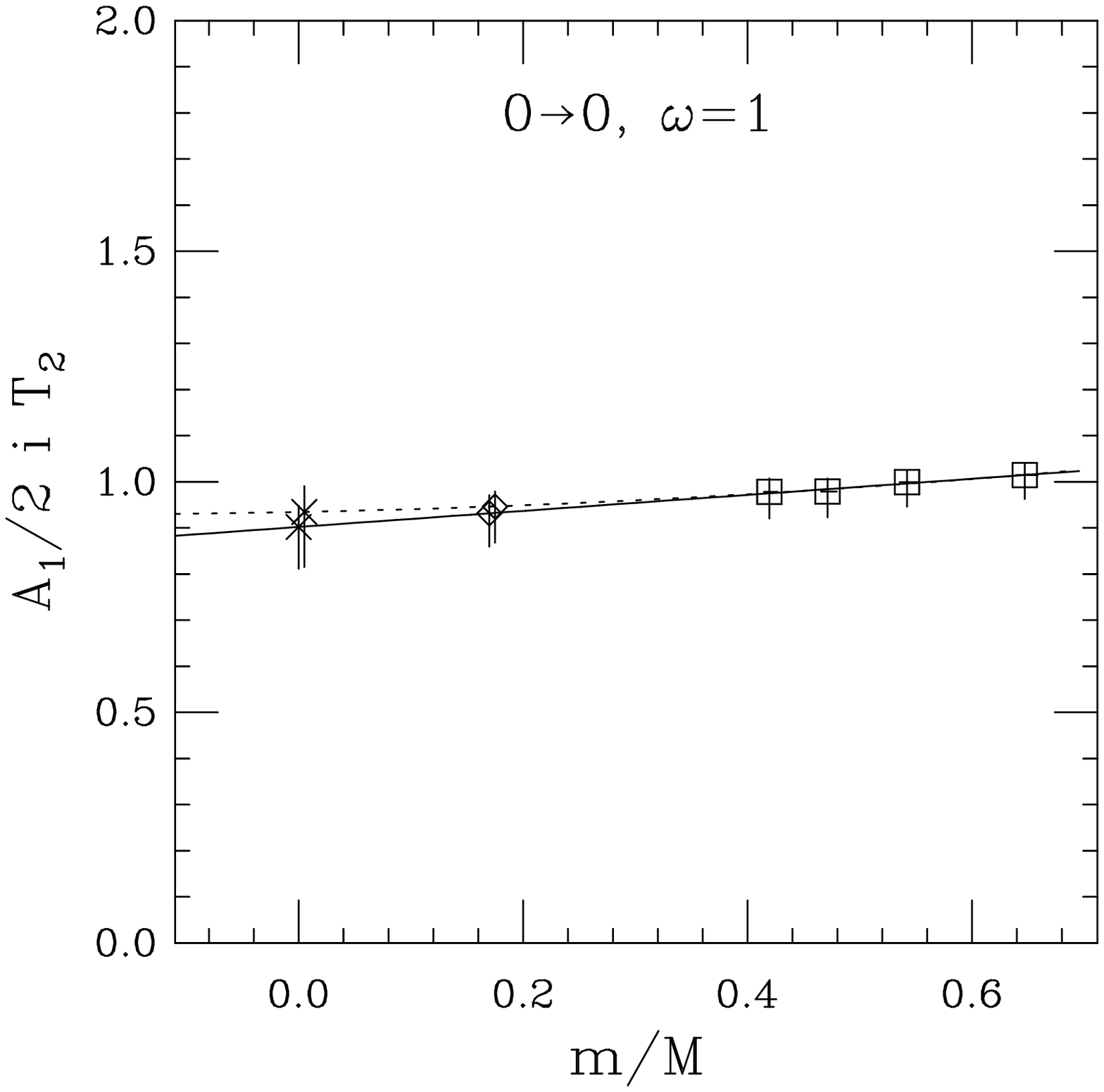}
\hfill\epsfxsize=0.48\hsize
\epsffile[37 37 510 510]{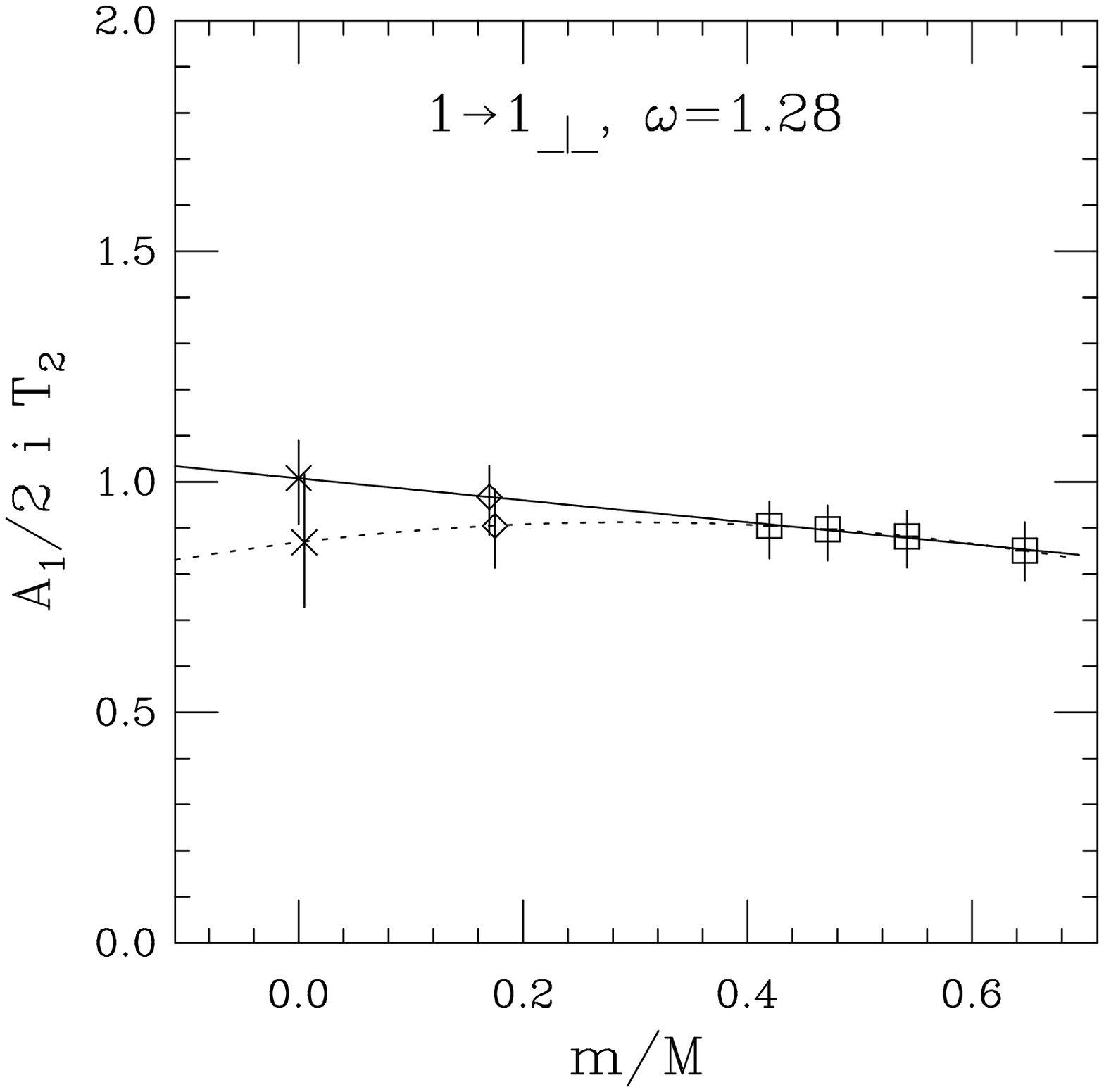}}
}
\caption[]{$V/2T_1$ and $A_1/2iT_2$ as a function of the ratio of the
light vector meson mass over the heavy-light pseudoscalar meson mass,
for different heavy quark masses (squares) and various momentum
channels. For the $0\to0$ channel $\w$ is $1$ exactly. For the
remaining channels, the value of $\w$ shown in each plot has an error
of less than $1$ in the last digit. We show both linear (solid) and
quadratic (dashed) extrapolations in $1/M$. The extrapolations to the
$B$ scale and the infinite mass limit are indicated by diamonds and
crosses respectively.}
\label{fig:hqs1overM}
\end{figure}

\begin{figure}
\vbox{\offinterlineskip
\hbox to\hsize{\hfill
\epsfysize=0.43\hsize
\epsffile[35 35 232 510]{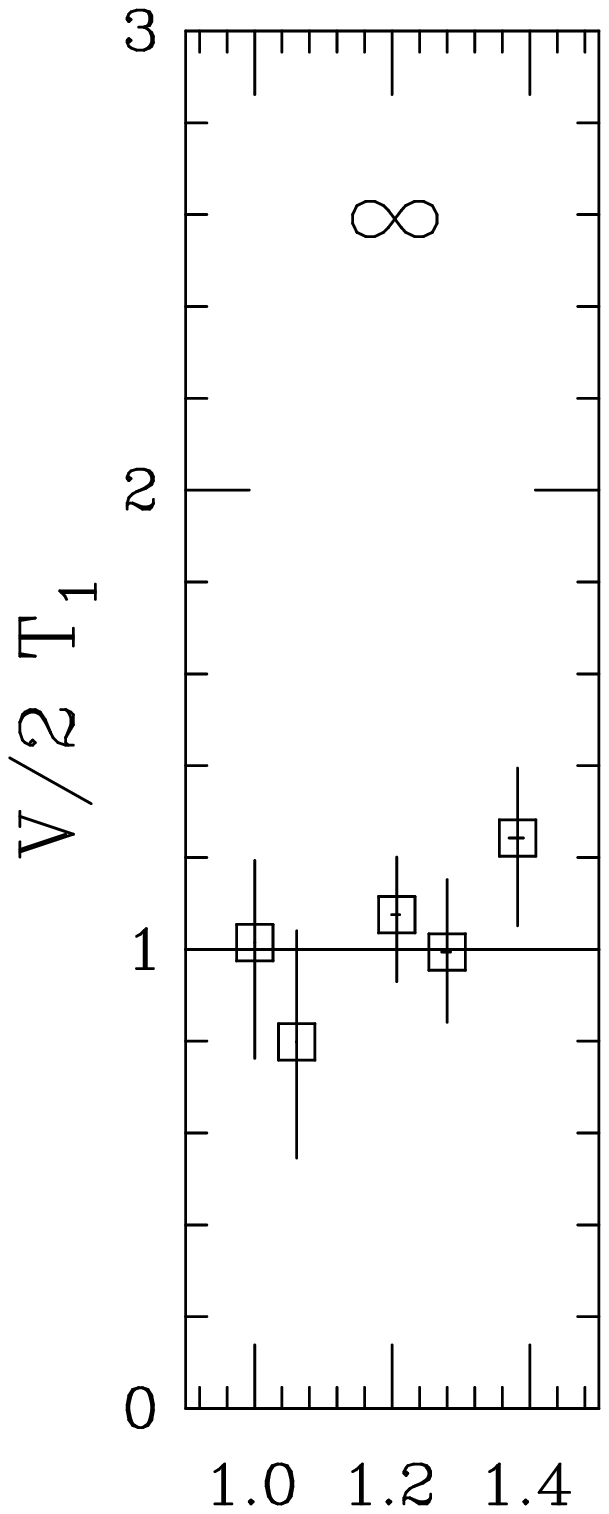}
\epsfysize=0.43\hsize
\epsffile[73 35 232 510]{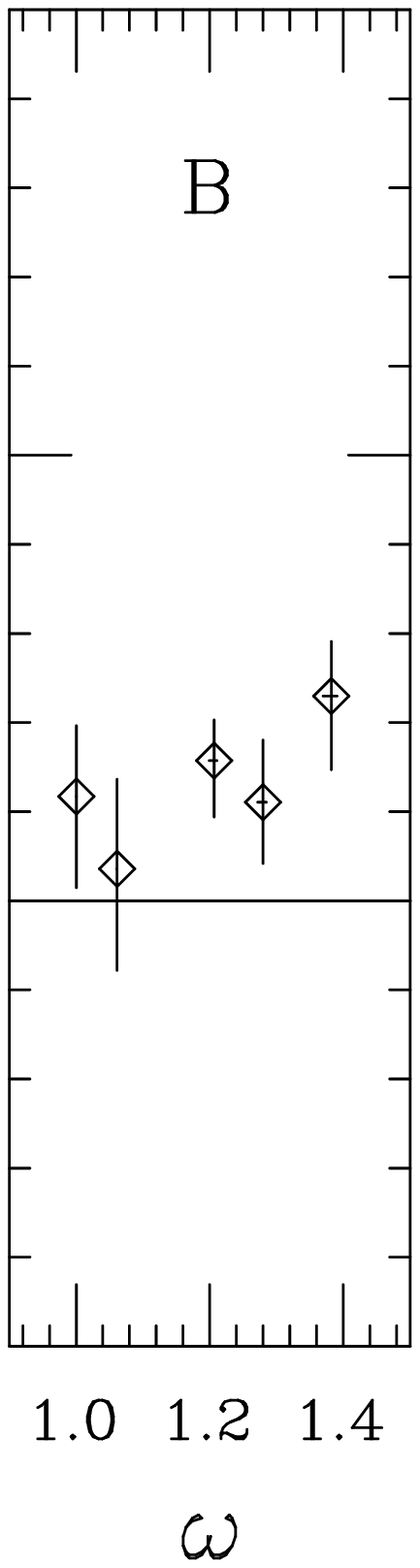}
\epsfysize=0.43\hsize
\epsffile[73 35 232 510]{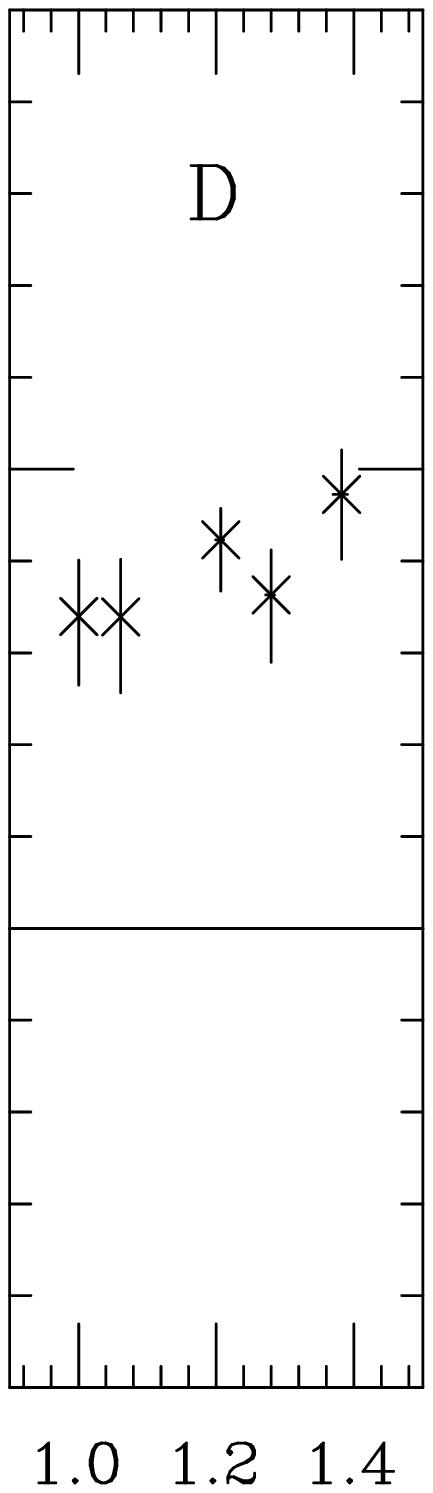}
\hfill}\kern1em
\hbox to\hsize{\hfill
\epsfysize=0.43\hsize
\epsffile[35 35 232 510]{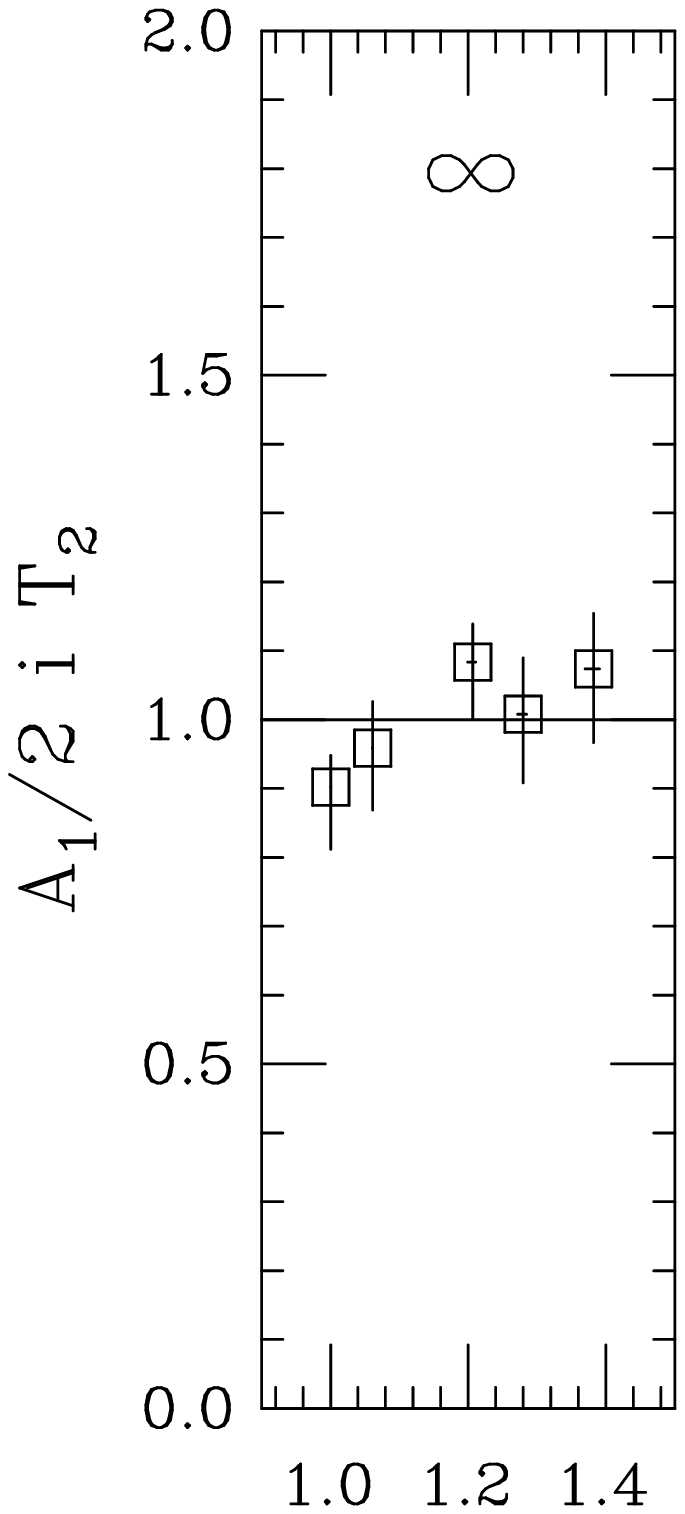}
\epsfysize=0.43\hsize
\epsffile[73 35 232 510]{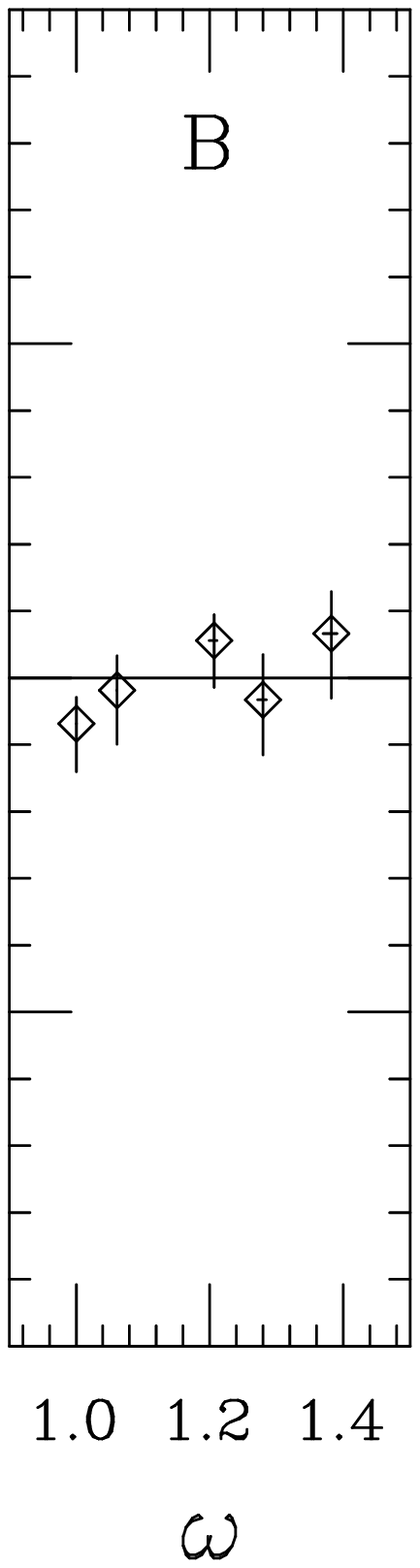}
\epsfysize=0.43\hsize
\epsffile[73 35 232 510]{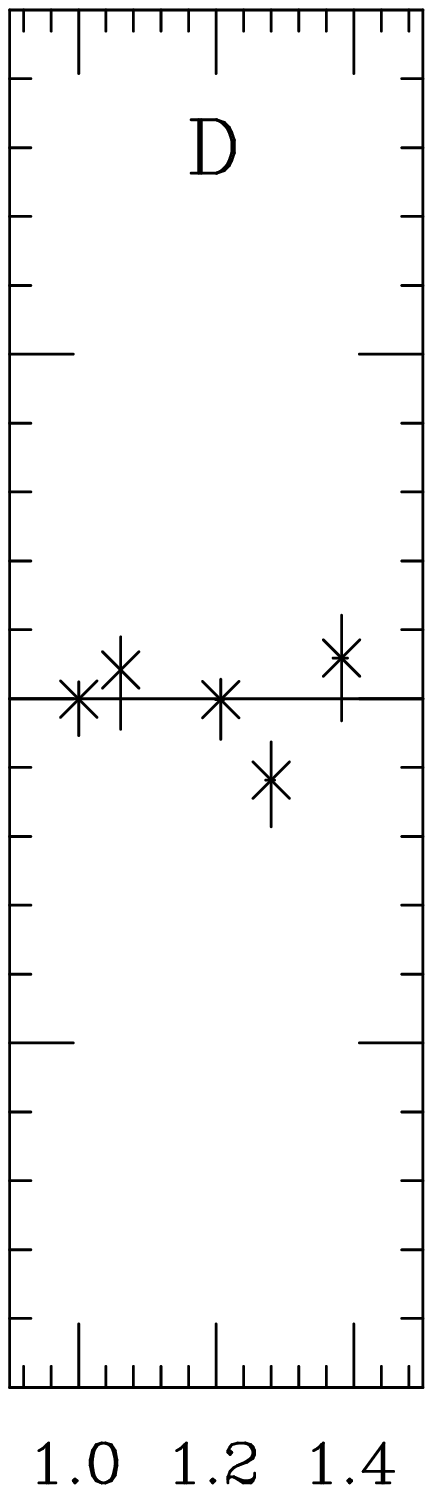}\hfill}}
\caption[]{Ratios $V/2T_1$ and $A_1/2iT_2$ for five values of $\w$
at three different heavy-light pseudoscalar masses, around the $D$
mass (crosses), around the $B$ mass (diamonds) and in the infinite
mass limit (squares). The horizontal solid line denotes the HQS
prediction in the infinite mass limit.}
\label{fig:hqsomega}
\end{figure}

In all cases the ratios $V/2T_1$ and $A_1/2iT_2$ are consistent with
unity in the heavy quark mass limit, as predicted by HQS, although for
certain channels the ratio $V/2T_1$ departs from $1$ by up to 75\% at
the charm scale. This constitutes a non-trivial test of HQS
predictions. These results are illustrated in
figures~\ref{fig:hqs1overM} and~\ref{fig:hqsomega}.

In figure~\ref{fig:hqs1overM} we show the ratios $V/2T_1$ (for
momentum channels $0\to1$ and $1\to0$) and $A_1/2iT_2$ (for momentum
channels $0\to0$ and $1\to1_\perp$) for different pseudoscalar meson
masses, allowing for linear and quadratic $1/M$ corrections to the
infinite mass limit predictions of equation~(\ref{eq:VT1andAT2}). The
linear and quadratic extrapolations in $1/M$ agree within errors for
the extrapolated values, with the exception of three cases of the
infinite mass limit points (one of which is shown in
figure~\ref{fig:hqs1overM}). We trust the linear extrapolations more
than the quadratic ones: we are fitting to four points only in a
limited region of inverse heavy-light pseudoscalar mass, and quadratic
fits can amplify accidental quadratic effects in the fitted points.
All results in the remainder of this section will refer to linear
extrapolations.

Figure~\ref{fig:hqsomega} shows the ratios $V/2T_1$ and $A_1/2iT_2$
for five values of $\w$ at three different heavy-light pseudoscalar
masses, around the $D$ mass, around the $B$ mass and in the infinite
mass limit. The results around the $D$ mass are our measured values at
$\kappa_h = 0.129$ (which corresponds very closely to the charm
quark). The $B$ mass and infinite mass limit results are
extrapolations. The HQS predictions of equation~(\ref{eq:VT1andAT2})
are well satisfied for both ratios in the infinite mass limit. The
ratio $V/2T_1$ shows large $1/M$ corrections of the order of 75\% at
the $D$ scale and 20\% at the $B$ scale. Corrections of about 30\% at
the $D$ scale and about 10\% at the $B$ scale were previously observed
for the pseudoscalar meson decay constant $f_P$~\cite{ukqcd:fb} and
for the form factor $h_V$ in heavy-to-heavy $0^-\to1^-$ semileptonic
decays~\cite{ukqcd:lpllat94}\footnote{Short distance corrections have
been accounted for in the form factors $h_V$, $h_+$ and $h_{A_1}$.}.
In contrast, the ratio $A_1/2iT_2$ exhibits small $1/M$ corrections
even at the $D$ meson scale. A similar situation occurs for the form
factors $h_+$ and $h_{A_1}$ in heavy-to-heavy $0^-\to0^-,1^-$
semileptonic decays~\cite{ukqcd:btodstar,ukqcd:lpllat94}\footnote{Note
that $h_+$ and $h_{A_1}$ are protected from $1/M$ corrections at zero
recoil by Luke's theorem~\protect\cite{luke} and the leading
corrections are of order $1/M^2$.}.

Inclusion of kinematic $1/M$ factors, as shown in
equations~(\ref{eq:VT1kinematic}) and (\ref{eq:AT2kinematic}) reduces
the size of the $1/M$ corrections by about a factor of two for the
relation between $T_1$ and $V$, giving results in agreement with those
in reference~\cite{sumrules:abs}, but has little effect on the $T_2$
and $A_1$ relation (note that the $T_2$--$A_1$ relation in
equation~(\ref{eq:AT2kinematic}) reduces to the result in
equation~(\ref{eq:VT1andAT2}) at $\w=1$).

\section{Cabibbo Suppressed Decays}

For this exploratory study we did not have a complete set of light
quark kappa values for every heavy-quark kappa value, as discussed in
section~\ref{sec:latticedetails}. This prevented us,
in~\cite{ukqcd:hlff}, from performing a reliable chiral extrapolation
to obtain results for the semileptonic decay $\btopi$. Here we are
concerned with semileptonic decays with a light vector meson in the
final state: in contrast to the case of a pion, which is a
pseudo-Goldstone boson, we do not expect a sizable effect from the
chiral extrapolation.

\begin{figure}
\hbox to\hsize{\epsfxsize=0.48\hsize
\epsffile[29 36 529 510]{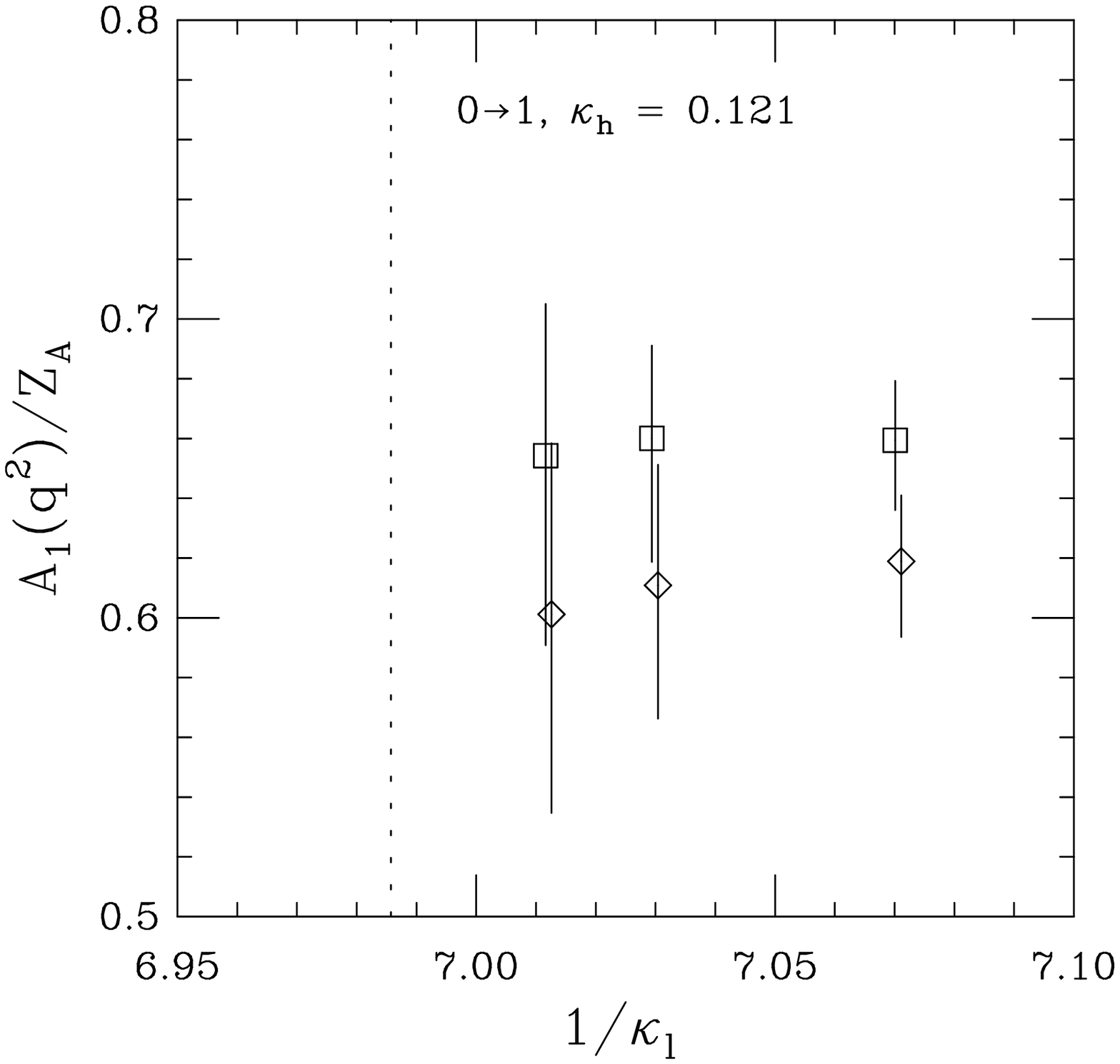}
\hfill\epsfxsize=0.48\hsize
\epsffile[29 36 529 510]{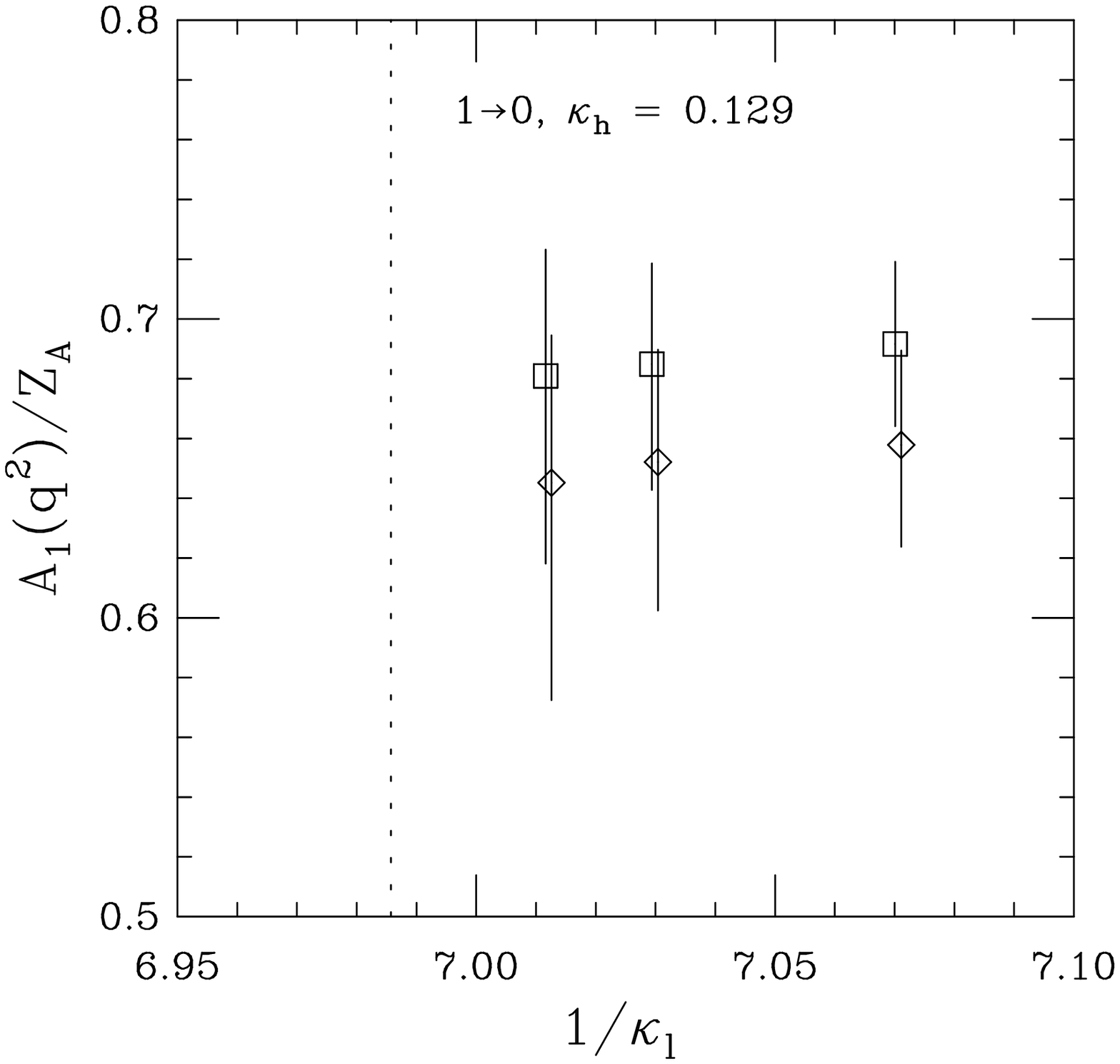}}
\caption[]{$A_1/Z_A$ for two different momentum channels and six
different combinations of spectator and active quark masses as a
function of the inverse spectator quark kappa value. Squares denote
$\kappa_a = 0.14144$ and diamonds denote $\kappa_a = 0.14226$. The
diamond points are slightly displaced horizontally for clarity. The
heavy kappa values are $\kappa_h = 0.121$ for the $0\to1$ channel and
$0.129$ for the $1\to0$ channel. The vertical dashed line marks the
chiral limit for the light spectator quark.}
\label{fig:sixkappas}
\end{figure}

Indeed, in figure~\ref{fig:sixkappas} we show the dependence of the
form factor $A_1$ on the active and spectator light-quark masses, for
two different combinations of momentum channel and heavy quark mass,
$\kappa_h = 0.121$ and $0.129$, for which six combinations of light
quark kappas were computed. It can be seen that in both cases $A_1$
remains practically constant as the light spectator quark mass
decreases (horizontal movement on the plots) but there is a dependence
on the active quark mass (vertical movement). Note that the value of
$\w$, and hence $q^2$, depends on $\kappa_l$ as well as $\kappa_a$,
and therefore the results of figure~\ref{fig:sixkappas} indicate that
$A_1^{\kappa_l,\kappa_a}\big(\w(\kappa_l,\kappa_a)\big)$ is nearly
independent of $\kappa_l$, in which case,
\begin{equation}\label{eq:spec-indepce}
A_1^{\kappa_{\rm crit},\kappa_{\rm crit}}
  \big(\w(\kappa_{\rm crit},\kappa_{\rm crit})\big) \approx
     A_1^{\kappa_l{=}0.14144,\kappa_{\rm crit}}
       \big(\w(\kappa_l{=}0.14144,\kappa_{\rm crit})\big).
\end{equation}
Similar independence of
the spectator mass was found in the study of the form factor $T_2$ in
the radiative decay $\btokstargamma$~\cite{ukqcd:bsg2}. As we have
shown in the previous section, $A_1$ is equal to $2iT_2$ to good
approximation for different heavy quark masses and different momentum
channels and therefore the results of~\cite{ukqcd:bsg2} give us
further evidence of the spectator mass independence of $A_1$.

\begin{figure}
\hbox to\hsize{\epsfxsize=0.48\hsize
\epsffile[29 36 529 510]{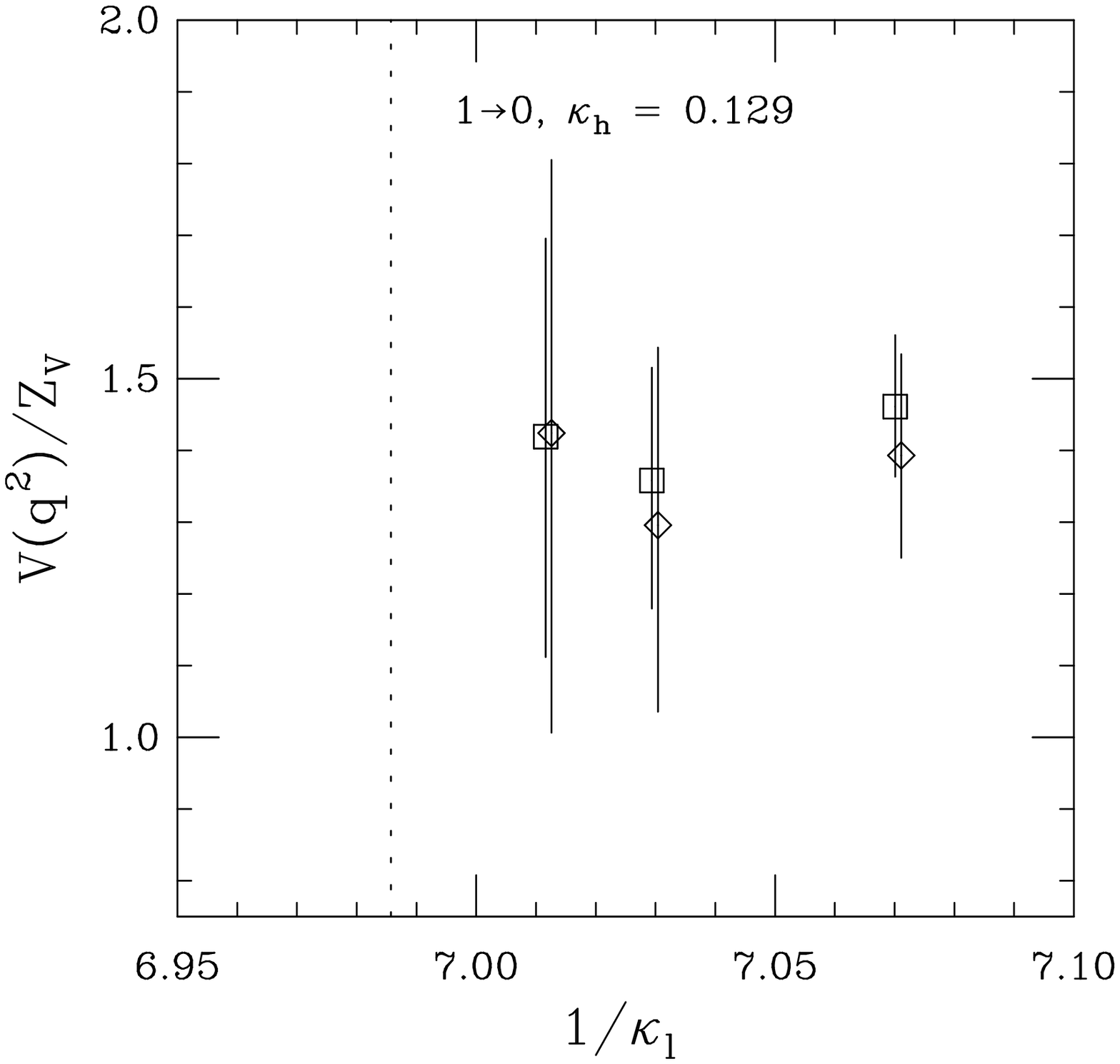}
\hfill\epsfxsize=0.48\hsize
\epsffile[29 36 529 510]{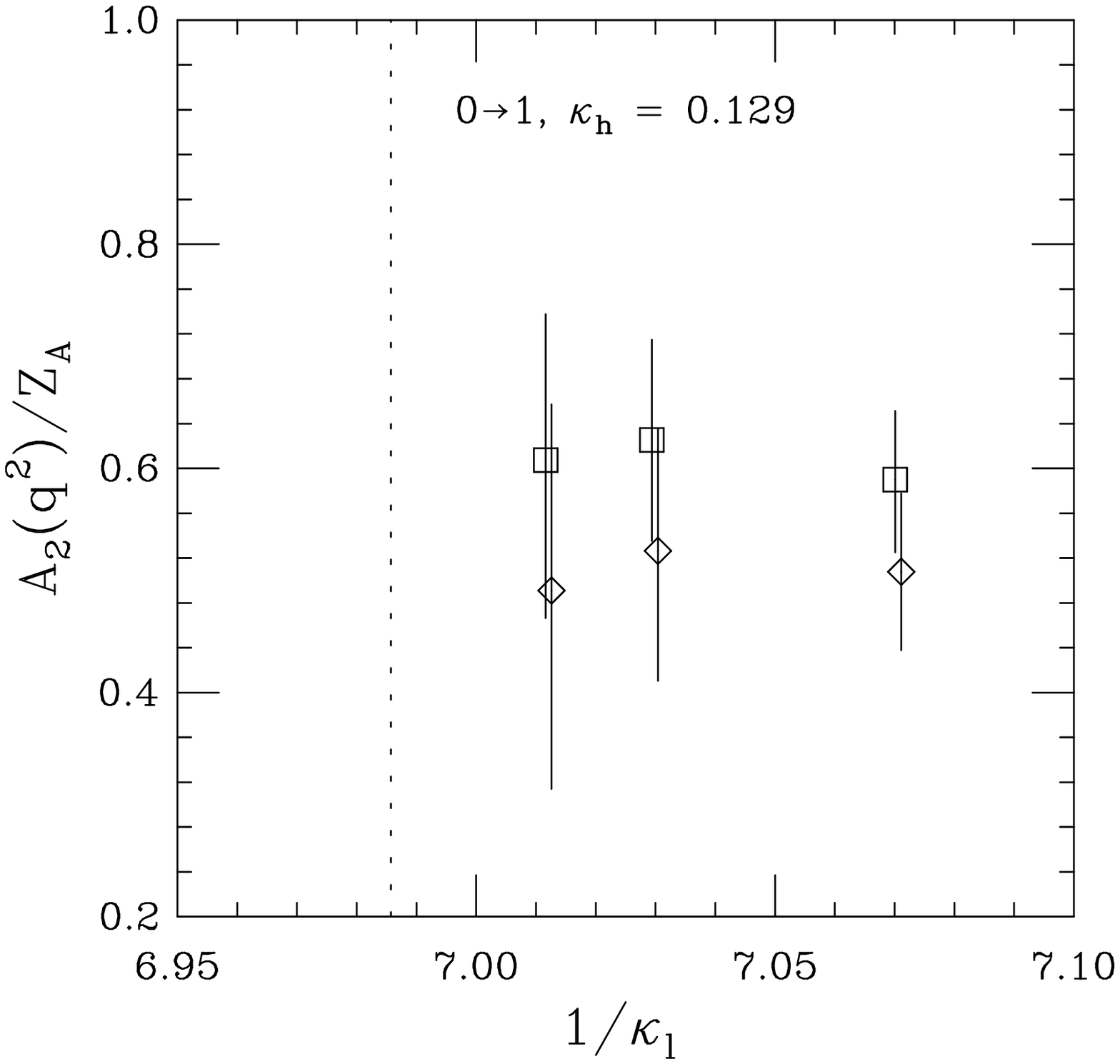}}
\caption[]{$V(q^2_{1\to0})/Z_V$ and $A_2(q^2_{0\to1})/Z_A$ for two
different momentum channels and six different combinations of
spectator and active quark masses as a function of the inverse
spectator quark kappa value. Squares denote $\kappa_a = 0.14144$ and
diamonds denote $\kappa_a = 0.14226$. The diamond points are slightly
displaced horizontally for clarity. The heavy kappa value is $\kappa_h
= 0.129$ in both cases. The vertical dashed line marks the chiral
limit for the light spectator quark.}
\label{fig:sixkappas-V-A2}
\end{figure}

The situation is less clear for the form factors $A_2$ and $V$ owing
to the larger statistical errors. $A_2$ and $V$ appear to follow the
same pattern as $A_1$, but we cannot dismiss a possible mild
dependence on the light spectator mass. In
figure~\ref{fig:sixkappas-V-A2} we show two examples of the dependence
of the form factors $V$ and $A_2$ on the active and spectator light
quark masses.

We are going to assume the form factors are independent of the light
spectator mass, and therefore we will extrapolate to the chiral limit
only for the active light quark, as was suggested for $A_1$ in
equation~(\ref{eq:spec-indepce}). This will allow us to use results
for all four heavy quark masses to guide our extrapolations to the $B$
mass. To quantify the systematic error induced we have compared, for
the two heavy kappa values for which we have six combinations of light
quark masses, the chiral-extrapolated form factor values for each
momentum channel using two procedures:
\begin{enumerate}
\item Using all six data points we can extrapolate to the chiral limit
for both the spectator and active quarks (see
reference~\cite{ukqcd:dtok}).
\item Fixing the spectator kappa value to $0.14144$ and assuming
complete independence of the form factors on the spectator quark mass,
we perform the chiral extrapolation for the active quark only, as
indicated in equation~(\ref{eq:spec-indepce}) for $A_1$.
\end{enumerate}
For $A_1$ the two procedures give the same results within 5\% except
for those channels involving a momentum of $\sqrt2$ in lattice units
for the light vector meson, where the variation can be 10\%.  These
errors are comparable with the statistical ones. For procedure 1
above, the chiral extrapolation for the $\sqrt2$ channels is less
reliable than for channels with lower light meson momentum, because
the statistical fluctuations for three-point functions with the
lightest light-quark masses ($\kappa_l = 0.14226, 0.14262$) are
larger.  These fluctuations can account for an important part of the
10\% variation observed between the two procedures. We have decided to
use 5\% to estimate, for all momentum channels, the systematic error
on $A_1$ induced by this assumption of independence of the light
spectator quark mass.

The statistical fluctuations are much larger for $A_2$ and $V$, making
it difficult to check the difference between the results from the two
chiral extrapolation procedures above. We will adopt a conservative
position and will admit a 20\% systematic error in $A_2$ and a 10\%
error in $V$, based on the maximum discrepancies observed between the
two procedures. These discrepancies are within the statistical error
bars.

\begin{figure}
\vbox{%
\hbox to\hsize{\epsfxsize=0.48\hsize
\epsffile[30 33 511 510]{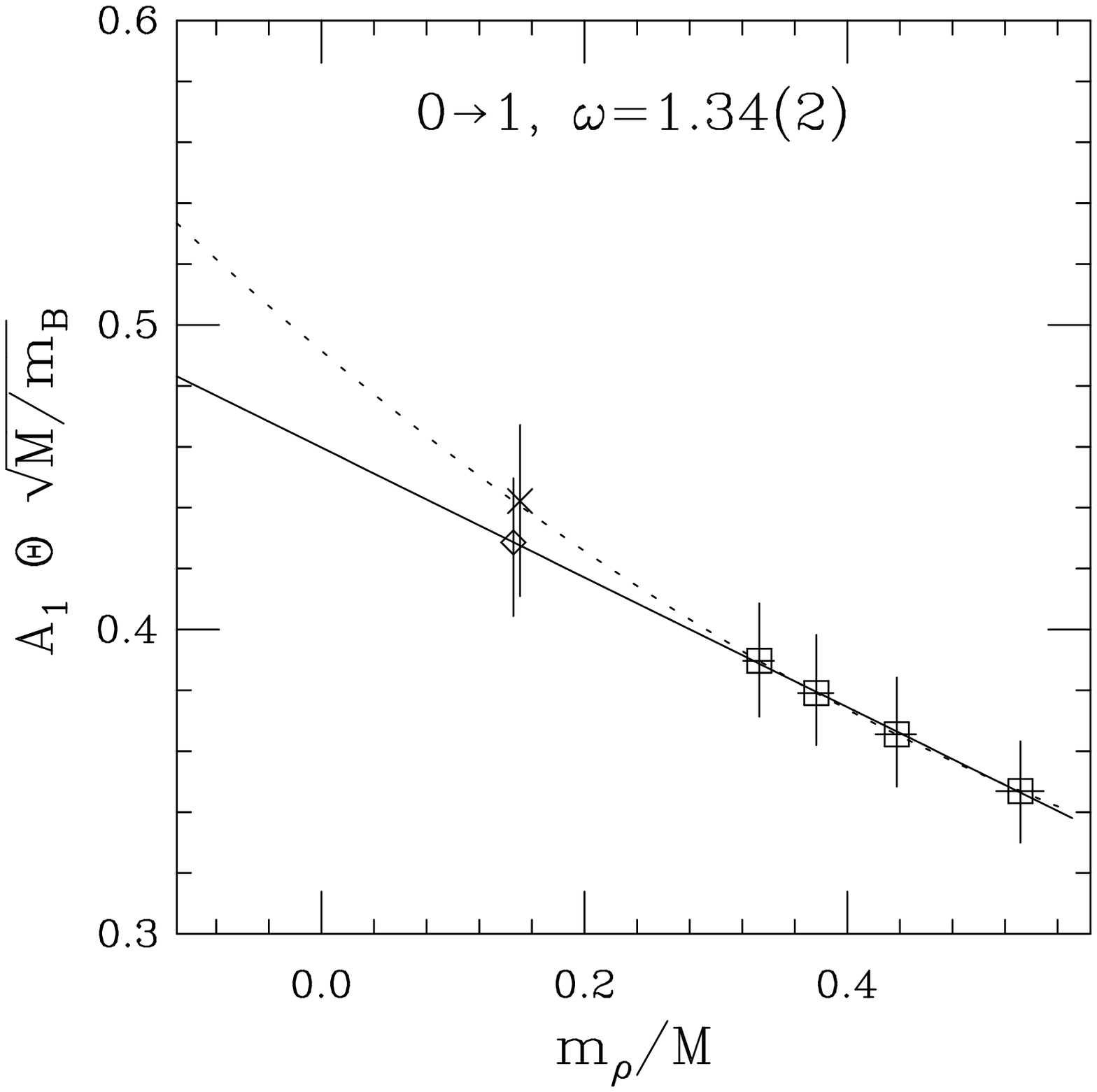}
\hfill\epsfxsize=0.48\hsize
\epsffile[30 33 511 510]{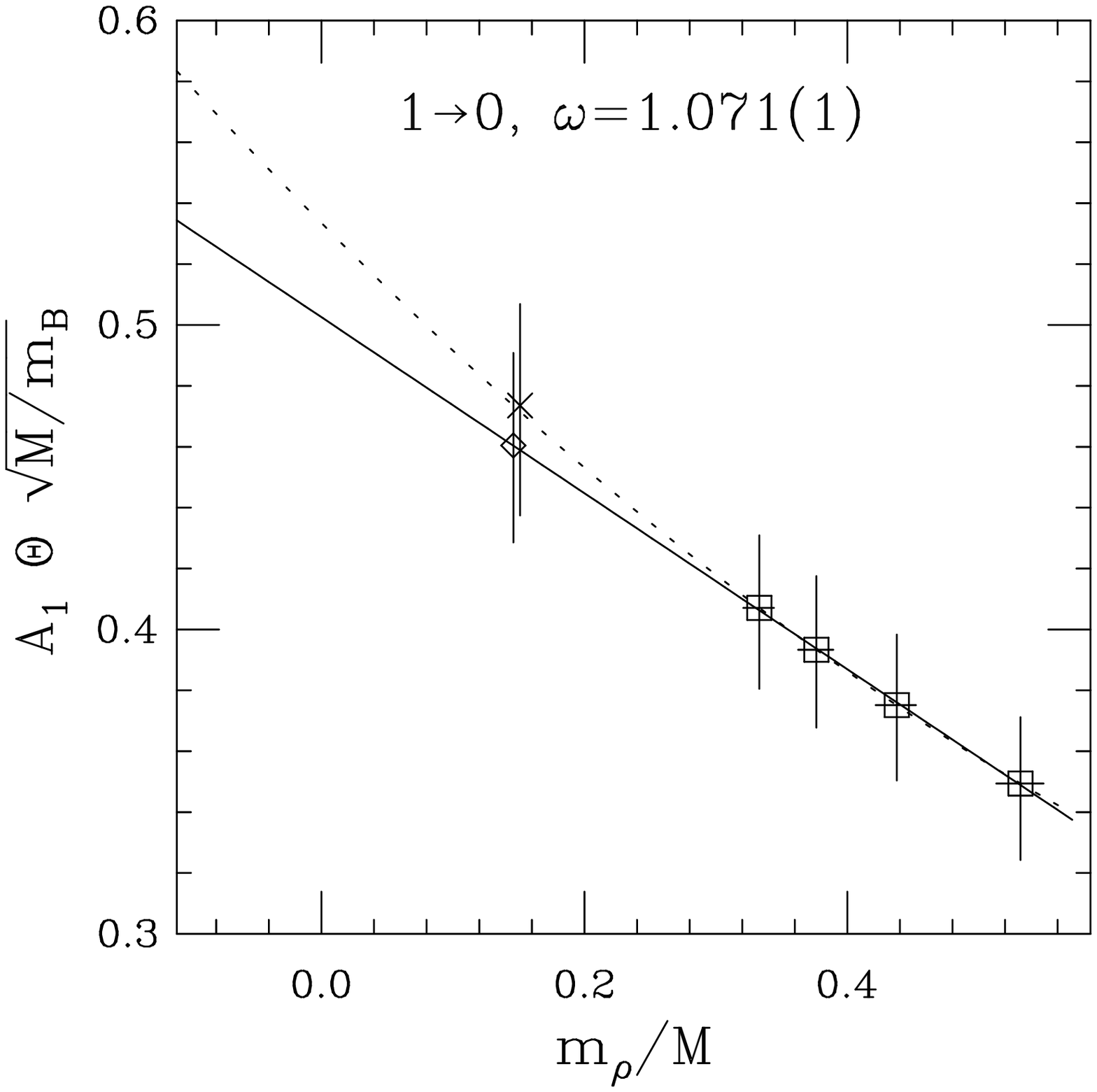}}
\kern1em
\hbox to\hsize{\epsfxsize=0.48\hsize
\epsffile[30 33 511 510]{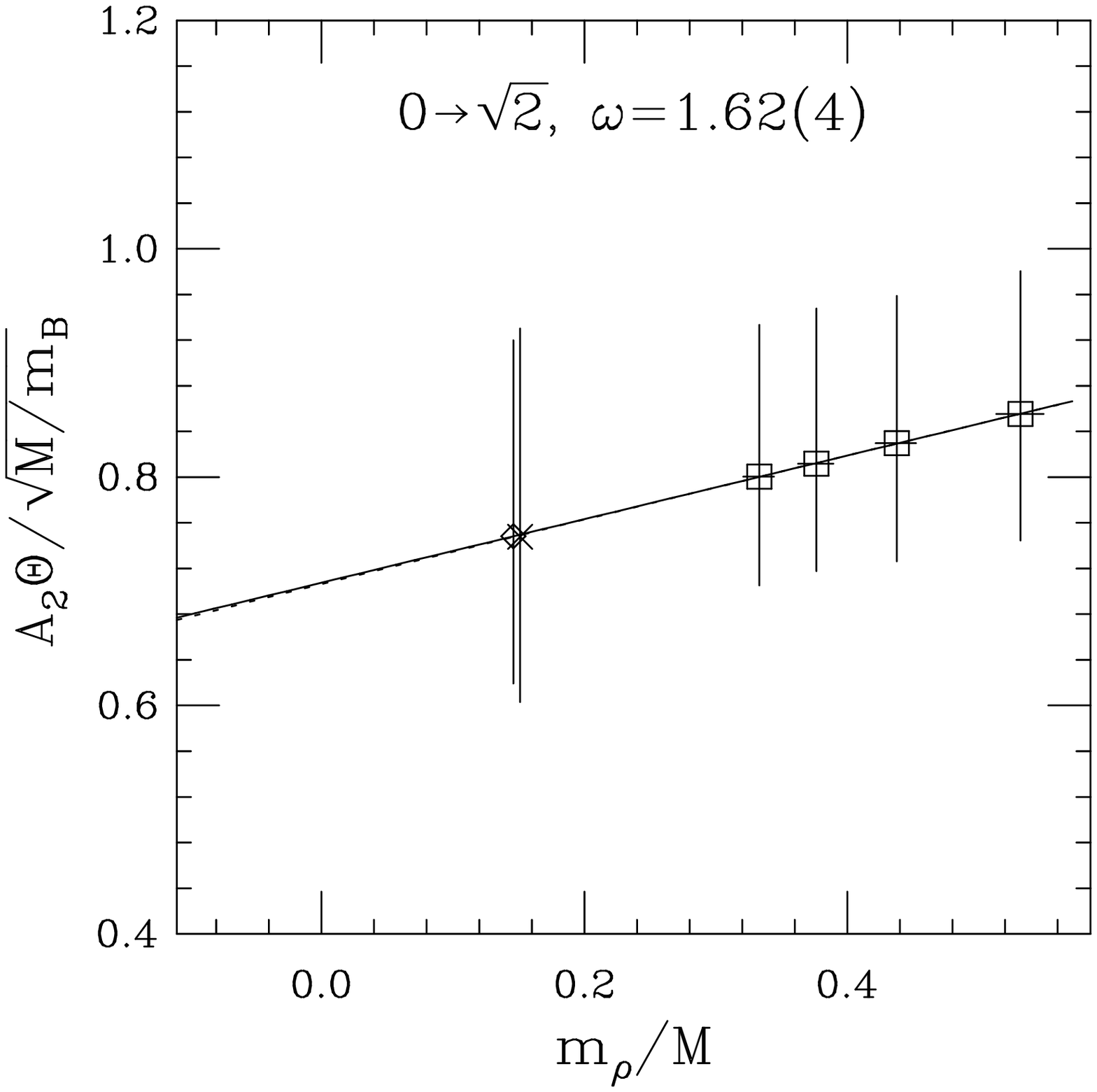}
\hfill\epsfxsize=0.48\hsize
\epsffile[30 33 511 510]{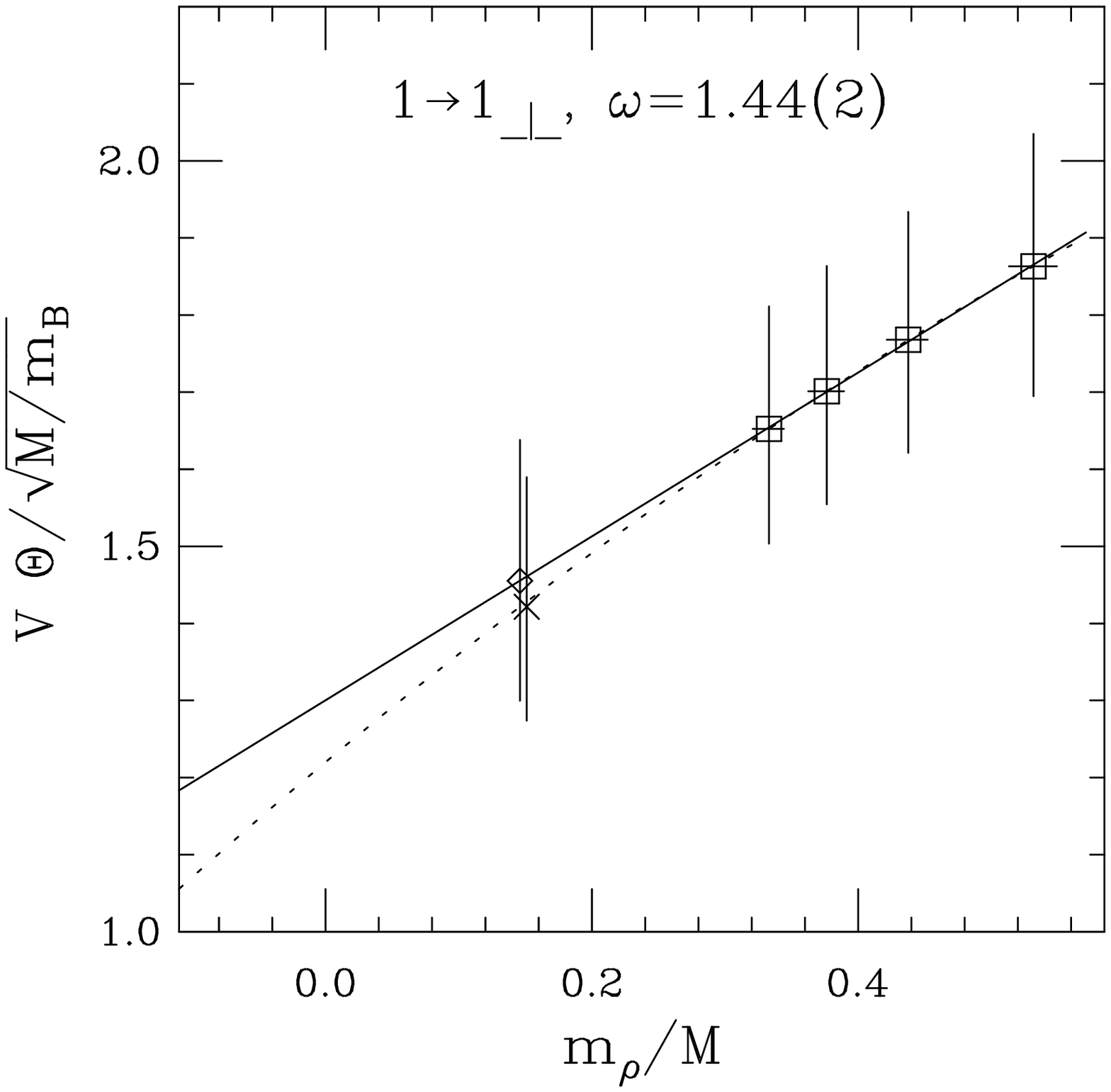}}}
\caption[]{Linear (solid line) and quadratic (dashed line) $1/M$
extrapolations of $A_1$, $A_2$ and $V$ in selected momentum channels
as functions of $m_\rho/M$. Squares denote measured points. Diamonds
and crosses mark the extrapolations to the $B$ scale for the linear
and quadratic fits respectively.}
\label{fig:A1A2V-1overM-extrapolation}
\end{figure}

In figure~\ref{fig:A1A2V-1overM-extrapolation} we show extrapolations
to the $B$ mass scale for the form factors $A_1$, $A_2$ and $V$,
allowing for linear and quadratic $1/M$ corrections to the infinite
mass limit predictions of equation~(\ref{eq:btorho-infmass}). The
extrapolated values agree well for the linear and quadratic fits: we
will quote results only from the linear fits in the following, unless
stated otherwise. We have used six momentum channels: the same five
channels used in the previous section together with the channel
$1\to\sqrt2_\perp$. For $A_2$ and $V$ the momentum channels $0\to 0$
and $1\to0$ are extremely noisy and we cannot control the
extrapolations\footnote{The $0\to0$ momentum channel is not measured
directly for $A_2$ and $V$.}.  The measured values of the form
factors, extrapolated to the $B$ scale for different values of $q^2$
close to $\qsqmax$, are listed in table~\ref{tab:formfactorvalues}.
\def\er#1#2{$^{+#1}_{-#2}$}
\def\tstrut{\vrule height2.5ex width0pt depth0pt}
\begin{table}
\hbox to\hsize{\def\arraystretch{1.2}\hss\begin{tabular}{lcccc}
\hline
\tstrut
channel & $q^2/\gev^2$ & $A_1$ & $A_2$ & $V$ \\[0.5ex]
\hline
\tstrut
$0 \to 0$       & 20.3      & 0.46\er23 & ---      & ---      \\
$0 \to 1$       & 17.5\er22 & 0.43\er22 & 0.8\er22 & 1.6\er11 \\
$0 \to \sqrt2$  & 15.3\er33 & 0.39\er32 & 0.7\er21 & 1.2\er11 \\
$1 \to 0$       & 19.7\er11 & 0.46\er33 & ---      & ---      \\
$1 \to 1_\perp$ & 16.7\er22 & 0.38\er33 & 0.6\er33 & 1.5\er22 \\
$1 \to \sqrt2_\perp$ & 14.4\er33 & 0.39\er65 & 0.7\er32 & 1.4\er32 \\[0.5ex]
\hline
\end{tabular}\hss}
\caption[]{Values of the form factors for $\b2rho$ for different
values of $q^2$ close to $\qsqmax$. For the $0\to0$ channel, $\w$ is
$1$ exactly, so that $q^2$ is fixed up to (tiny) experimental errors
in the physical meson masses. For $A_2$ and $V$, the zero recoil
channel, $0\to0$ is not measured. In addition, the large statistical
errors in $A_2$ and $V$ for the $1\to0$ channel prevent a reliable
extrapolation in $1/M$. Quoted errors are purely statistical. Adding
systematic errors from spectator quark flavour symmetry breaking and
discretisation in quadrature produces a further 11\% error in $A_1$,
20\% in $A_2$ and 15\% in $V$.}
\label{tab:formfactorvalues}
\end{table}

\begin{figure}
\hbox to\hsize{\epsfysize=0.46\hsize
\epsffile[35 37 510 510]{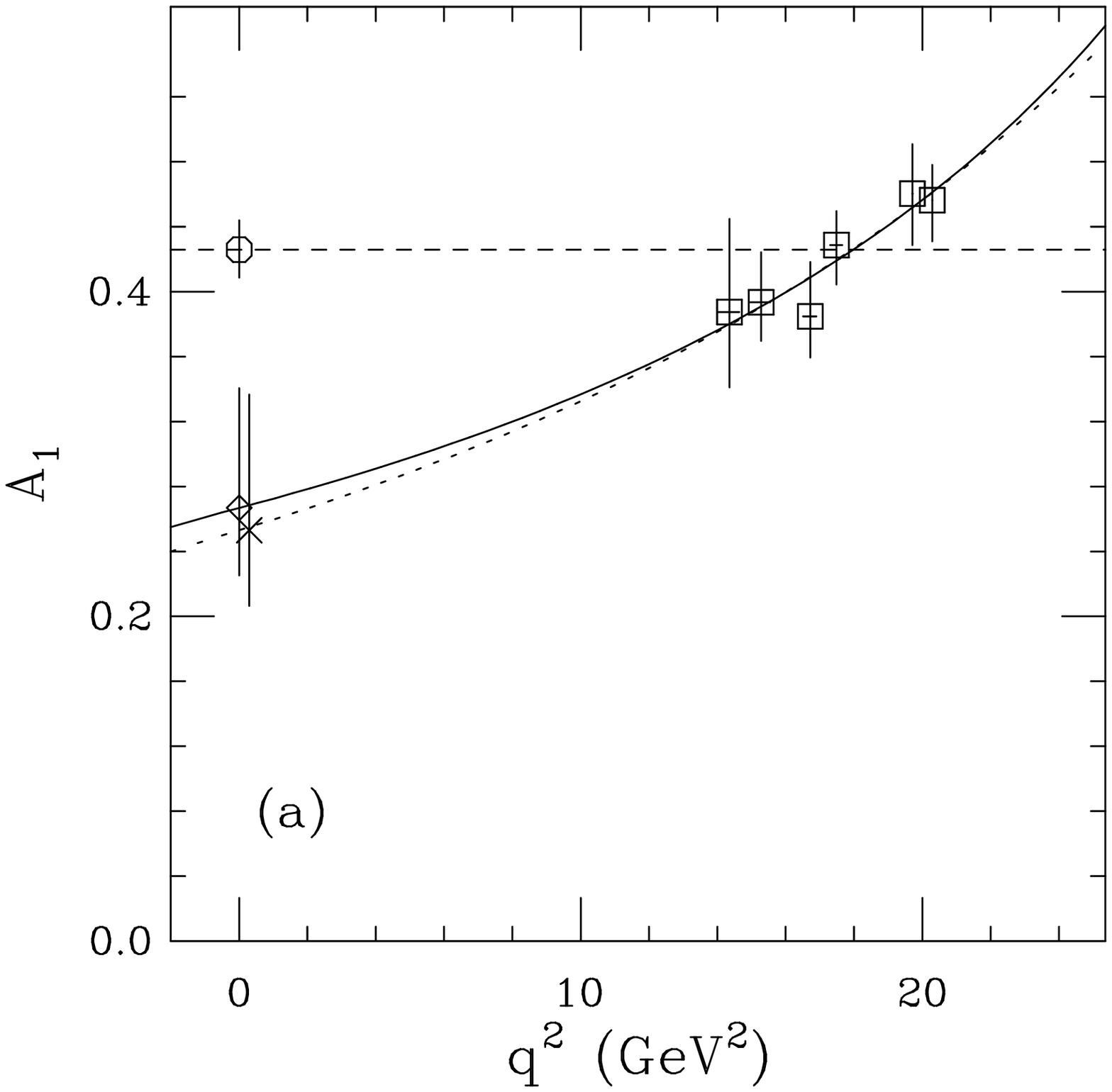}
\hfill\epsfysize=0.46\hsize
\epsffile[27 37 510 510]{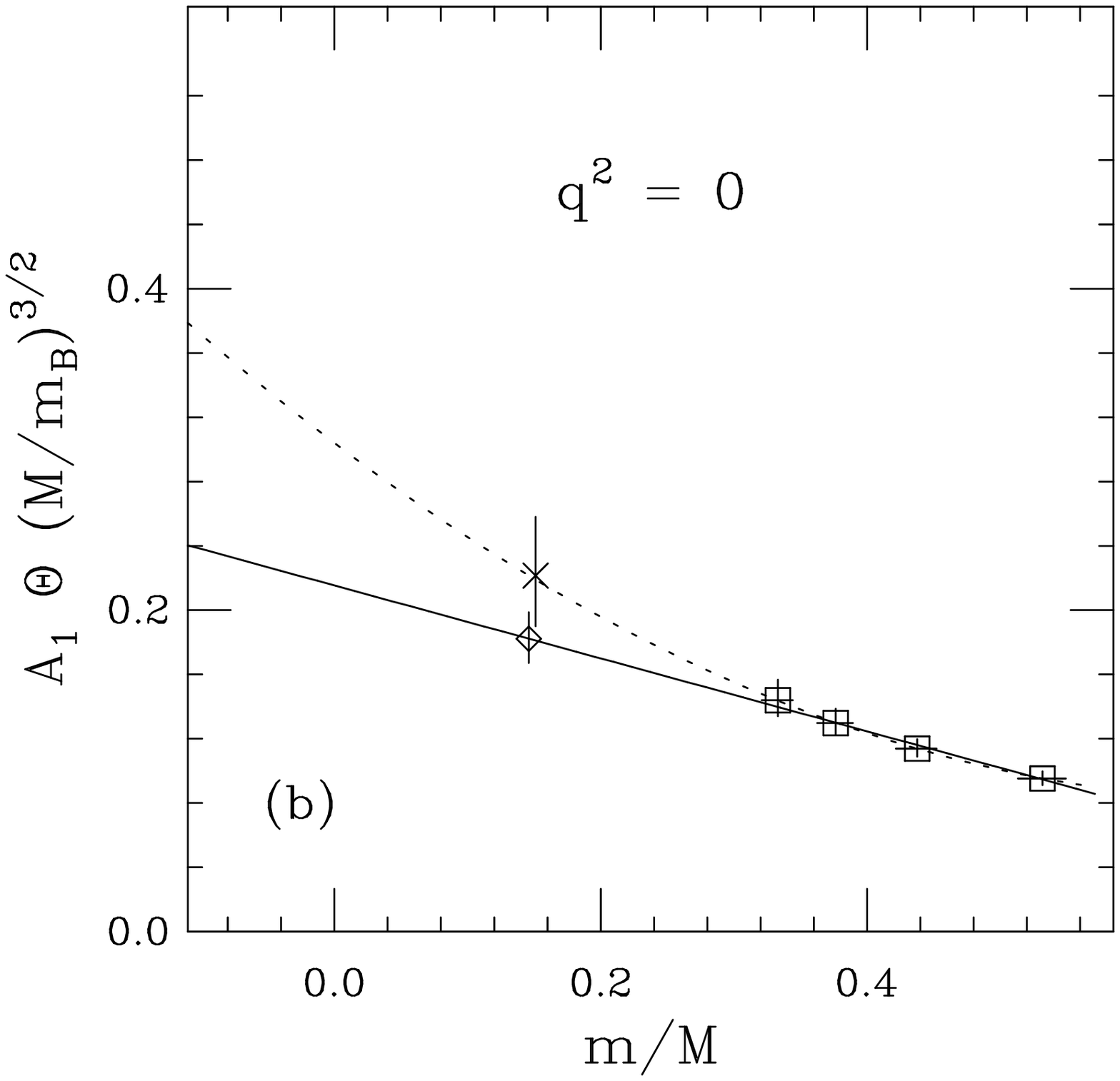}}
\caption[]{{\bf (a)}: form factor $A_1(q^2)$ for the decay $\b2rho$. Squares
are measured data extrapolated to the $B$ scale at fixed $\w$. The
three curves and points at $q^2=0$ have been obtained by fitting the
squares using the $q^2$ dependences from
equation~\protect(\ref{eq:qsqdepce}): constant (dashed line and
octagon), pole (solid line and diamond) and dipole (dotted line and
cross). The point at $q^2=0$ for the dipole fit has been displaced
slightly for clarity. {\bf (b)}: linear (solid) and quadratic
(dashed) extrapolations of $A_1(0)$ in $1/M$ to the $B$ scale
according to equation~\protect(\ref{eq:A1qsq0extrap}). Squares are
obtained from pole fits to the measured data for the different heavy
quark kappa values. The diamond and cross give the form factor at the
$B$ scale.}
\label{fig:A1mbqsq}
\end{figure}

In figure~\ref{fig:A1mbqsq}a we show $A_1(q^2)$ at the $B$ scale,
together with three fits to its $q^2$ dependence: \def\mstrut{\vrule
height4.5ex width0pt depth0pt}
\begin{equation}\label{eq:qsqdepce}
A_1(q^2) = \cases{A_1(0)&constant,\cr
         \displaystyle\mstrut
            {A_1(0)\over 1 - q^2/m_{\rm pole}^2}&pole,\cr
         \displaystyle\mstrut
            {A_1(0)\over(1-q^2/m_{\rm dipole}^2)^2}&dipole.\cr}
\end{equation}
Table~\ref{tab:A1mbqsq} shows the fit parameters for each $q^2$
dependence, with the corresponding $\chi^2$ per degree of freedom.
\begin{table}
\hbox to\hsize{\def\arraystretch{1.2}\hss\begin{tabular}{lcccc}
\hline
\tstrut
fit type & $A_1(0)$ & $m_{\rm pole}/\gev$ & $m_{\rm dipole}/\gev$ &
     $\chi^2/{\rm dof}$ \\[0.5ex]
\hline
\tstrut
constant & 0.43\er22 & ---    & ---    & 1.3 \\
pole     & 0.27\er74 & 7\er21 & ---    & 0.2 \\
dipole   & 0.25\er85 & ---    & 9\er41 & 0.2 \\
\hline
\end{tabular}\hss}
\caption[]{Fit parameters for different $q^2$ dependences for the form
factor $A_1(q^2)$ at the $B$ scale. Quoted errors are statistical only.}
\label{tab:A1mbqsq}
\end{table}

The figure and table indicate that constant-in-$q^2$ behaviour for
$A_1$ does not fit the data well. This feature is even more evident in
lattice studies at the $D$ scale, where fits to constant behaviour are
poor~\cite{ukqcd:dtok}%
--\cite{bks:dtok}, \cite{elc:DandB,ape:DandB}.  Our
data favour a pole type behaviour with a pole mass in the expected
range for a $1^+$ $b \bar u$ resonance~\cite{ape:DandB}. Dipole (and
in general higher powers: tripole, \ldots) and pole fits are hardly
distinguishable in the physical range $0 \leq q^2 \leq \qsqmax$. In a
dipole fit, the mass parameter $m_{\rm dipole}$ is roughly given by
$m_{\rm dipole} \approx \sqrt2 m_{\rm pole}$ and hence pole and dipole
fits agree in the limited range of values of $q^2$, well below $m_{\rm
pole}^2$, explored in figure~\ref{fig:A1mbqsq}a. An alternative
procedure is to use pole fits for extracting $A_1(\kappa_h;q^2{=}0)$
from our measured data, at different heavy quark masses around that of
the charm and then to extrapolate $A_1(0)$ in $1/M$ to the $B$ scale,
as shown in figure~\ref{fig:A1mbqsq}b, by
using~\cite{ukqcd:bsg2,ukqcd:hlff,ape:bsg}
\begin{equation}\label{eq:A1qsq0extrap}
A_1(0) \Theta M^{3/2} =
  {\rm const}(1 + \gamma/M + \delta/M^2 + \cdots).
\end{equation}
This gives a result for $A_1(0)$ at the $B$ scale (using linear
or quadratic extrapolations in $1/M$) of,
\begin{equation}
A_1(q^2{=}0;m_B) = \cases{0.18\pm0.02 & linear\cr
                          0.22^{+4}_{-3} & quadratic\cr}.
\end{equation}
The quoted errors are purely statistical. A further 11\% systematic
error should be added, obtained by combining in quadrature the
systematic errors from the spectator quark flavour symmetry breaking
and discretisation effects. Incorporating the systematic error makes
this method of extracting $A_1(0)$ consistent within errors with the
method presented in table~\ref{tab:A1mbqsq}.

In conclusion, we find that $A_1(q^2)$ at the $B$ scale is fitted by a
single pole form, with the parameters given in
table~\ref{tab:A1mbqsq}. The errors in table~\ref{tab:A1mbqsq} are
statistical only: as mentioned above, a further 11\% systematic error
should be added. In table~\ref{tab:A1-zero-results} we compare our
results with other theoretical calculations: the agreement is
generally good.  Previous lattice results~\cite{elc:DandB,ape:DandB}
rely on the assumption of pole behaviour using a lattice determination
of the $b\bar u$ $1^+$ resonance ($\bar B_1$) mass and the value of
$A_1(q^2)$ at a {\em single} $q^2$ value. In contrast, in this paper,
we have tried to determine the $q^2$ dependence of $A_1$.
\begin{table}
\hbox to\hsize{\hfill
\begin{tabular}{lc}
\hline
\tstrut
Reference & $A_1(0)$ \\[2pt]
\hline
\tstrut
This work                  & 0.27\er74\er33 \\
ELC ``a''~\cite{elc:DandB} & $0.25\pm0.06$ \\
ELC ``b''~\cite{elc:DandB} & $0.22\pm0.05$ \\
APE ``a''~\cite{ape:DandB} & $0.29\pm0.16$ \\
APE ``b''~\cite{ape:DandB} & $0.24\pm0.12$ \\
Sum rules~\cite{sumrules:abs}  & $0.24\pm0.04$ \\
Sum rules~\cite{sumrules:ball} & $0.5\pm0.1$ \\
Sum rules~\cite{sumrules:narison} & $0.35\pm0.16$ \\
Quark model~\cite{gisw} & $0.05$ \\
Quark model~\cite{bsw} & $0.28$ \\
Quark model~\cite{fgm} & $0.26\pm0.03$ \\
HQS \& $\chi$PT~\cite{A1zero:casalbuoni} & $0.21$ \\[2pt]
\hline
\end{tabular}\hfill}
\caption[]{Values for $A_1(0)$ for $\b2rho$ from various theoretical
calculations. The second set of errors in our quoted value denotes our
estimate of systematic effects from the spectator quark flavour
symmetry breaking and discretisation.}
\label{tab:A1-zero-results}
\end{table}

Statistical and systematic uncertainties in the present study have
prevented us from extracting the $q^2$ dependence of the $A_2$ and $V$
form factors.

Much effort has recently been devoted by the lattice community to
determining the $q^2$ behaviour of $T_1$ and $T_2$ around the $B$
scale, without producing a definitive conclusion.  The $q^2$
dependence found here for $A_1$ and the results from the previous
section for the ratio $A_1/2iT_2$ may support a pole-type behaviour
for $T_2$ at least in a region around $\qsqmax$. Sum-rules
calculations~\cite{sumrules:abs,sumrules:ball,sumrules:narison} and a
recent calculation by B~Stech~\cite{stech} find that $V$ has a more
pronounced $q^2$ dependence than $A_1$, which is consistent with
having dipole type behaviour for $V$ and pole behaviour for $A_1$. In
the previous section we found that around the $B$ scale $T_1$ is
roughly equal to $V$. This supports a dipole behaviour for $T_1$ and
consequently further favours a pole behaviour for
$T_2$~\cite{ukqcd:bsg2,ukqcd:hlff,ape:bsg}.

\section{Extraction of $\vub$}

In this section we employ our lattice determination of the form
factors to calculate the differential decay rate $d\Gamma/dq^2$ for
the decay $\b2rho$ in the region near $\qsqmax$. Experimental data for
this region will enable a model-independent determination of $\vub$.

It has previously been suggested that determinations of the decay rate
for $\b2rho$ near $q^2=0$, combined with the experimentally-measured
branching fractions $B(\btokstargamma)$ and $B(b\to s\gamma)$ would
also allow a determination of $\vub$. However, one method relies on
three different measurements, $B(\btokstargamma)$, $B(b\to s\gamma)$
and $d\Gamma(\b2rho)/dq^2|_{q^2=0}$, together with the theoretical
determination of the ratio
$T^{\btokstargamma}_1(0)/A^{B\to\rho}_0(0)$~\cite{odonnell,santorelli}.
The evaluation of this ratio of form factors from lattice calculations
involves difficult and model-dependent extrapolations from the
vicinity of $\qsqmax$ to $q^2=0$.  Another method uses the ratio of
$\Gamma(\btokstargamma)$ and $\lim_{q^2\to 0} d\Gamma(\b2rho)/dE_\rho
dE_l$ evaluated on a curve where $q^2 = 4E_l(m_B - E_\rho -
E_l)$~\cite{bd}, which is independent of any hadronic form factor.
However, it relies on the validity of the leading order HQS relations
of equations~(\ref{eq:VT1kinematic}) and~(\ref{eq:AT2kinematic})
between the form factors for the radiative and semileptonic decays in
the region around $q^2=0$, far from the zero recoil point.
Furthermore, light flavour $SU(3)$ symmetry is used to relate the
$K^*$ and the $\rho$ mesons. Corrections to these approximations will
induce systematic errors in the determination of $\vub$. For example,
the sum rule calculation of the relevant form factors in
reference~\cite{sumrules:abs} finds about 7\% corrections to the
leading HQS relations and about 25\% corrections from $SU(3)$
breaking.

We propose to look at the region around $\qsqmax$ for the process
$\b2rho$, beyond the region of charm production which complicates the
experimental determination of $b\to u$ transitions at low $q^2$ (note
that $|V_{ub}/V_{cb}|^2$ determined from inclusive decays is expected
to be less than $0.01$). The methods referred to above, based on
determinations of the decay rate near $q^2 =0$, suffer from this
experimental difficulty in addition to possible theoretical
uncertainties.  Lattice techniques, in contrast, allow the form
factors to be measured directly near $\qsqmax$, using only a $1/M$
extrapolation motivated by HQS, and avoiding the problematic
model-dependent extrapolation to $q^2 = 0$. This model-dependence
currently plagues the lattice determination of $B(\btokstargamma)$
which requires knowledge of the form factors $T_1$ and/or $T_2$ at
$q^2=0$, where they are equal.

The differential decay rate for $\b2rho$ is given by
\begin{eqnarray}
\frac{d\Gamma(\b2rho)}{dq^2} &=&
\frac{G_F^2 \vub^2}{192 \pi^3 m_B^3}
 q^2 [\lambda(q^2)]^{\frac{1}{2}}
\nonumber \\
& &
\mbox{} \times \left(|H^{+}(q^2)|^2
+ |H^{-}(q^2)|^2 + |H^{0}(q^2)|^2 \right)  \label{eq:diffdecayrate}
\end{eqnarray}
where $\lambda(q^2)= (m_B^2 + m_{\rho}^2 -q^2)^2 -4m_B^2m_{\rho}^2$.
$H^0$ comes from the contribution of the longitudinally polarised
$\rho$ and is given by
\begin{equation}
H^0(q^2)=\frac{-1}{2m_{\rho}\sqrt{q^2}}
\Big\{ (m_B^2-m^2_{\rho}-q^2)(m_B+m_{\rho})A_1(q^2)
- \frac{4m^2_B|{\bf k}|^2}{m_B+m_{\rho}}A_2(q^2)\Big\},
\end{equation}
where $\bf k$ is the momentum of the $\rho$ in the $B$-meson rest frame.
$H^{\pm}$ correspond to the contribution of the transverse polarisations
of the vector meson and are given by
\begin{equation}
H^{\pm}(q^2)=-\Big\{(m_B+m_{\rho})A_1(q^2)
\mp  \frac{2m_B|{\bf k}|}{(m_B+m_{\rho})}V(q^2)\Big\}.
\end{equation}

\begin{figure}
\hbox to\hsize{\hfill\epsfxsize=0.8\hsize
\epsffile[38 45 510 370]{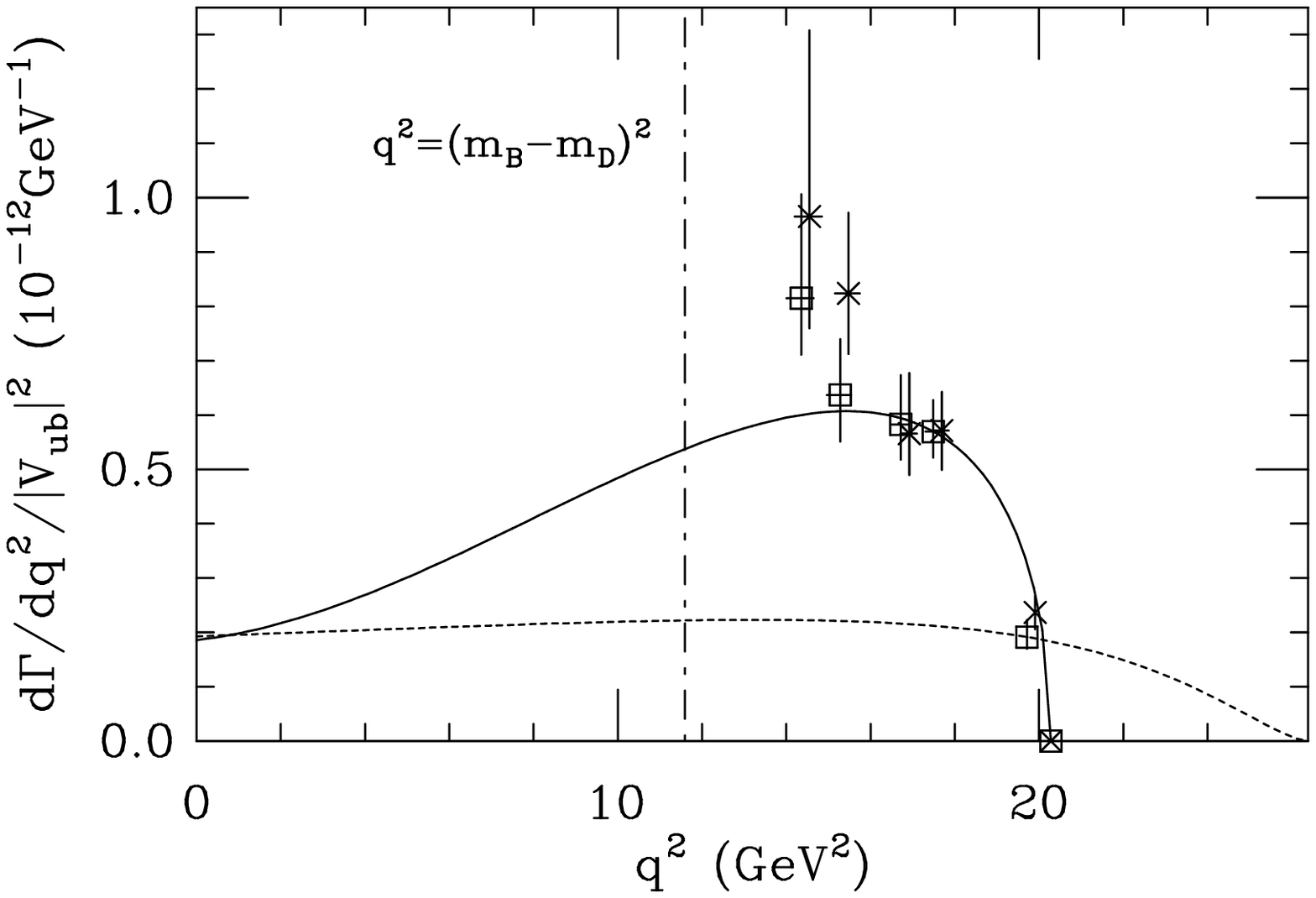}
\hfill}
\caption[]{Differential decay rates for $\b2rho$ and $\btopi$. The
data points and solid curve are for $\b2rho$: squares have
contributions from our measured values for $A_1$, $A_2$ and $V$,
crosses have $A_1$ only, and the crosses are offset slightly for
clarity. The dashed curve is for $\btopi$. The curves are
illustrative, as described in the text. The vertical dot-dashed line
marks the upper limit in $q^2$ for charm production.}
\label{fig:btorhovsbtopi}
\end{figure}

Looking for semileptonic decays beyond the charm production threshold
(determined by the semileptonic $\bar B \to D l \bar\nu_l$ decay) will
make sense only if there are enough events. In
figure~\ref{fig:btorhovsbtopi} we show the differential decay rates
$d\Gamma/dq^2$ for $\b2rho$ and $\btopi$. For $\b2rho$ we assume that
$A_1$, $A_2$ and $V$ are all given by single pole forms with a common
pole mass of $5.3\gev$ and a normalisation determined by our
measurements for the $0\to1$ momentum channel. For $\btopi$ we use the
results of reference~\cite{ukqcd:hlff} with a dipole form for the form
factor $f^+$ with $f^+(0) = 0.24$ and a mass parameter $m_{f^+} =
5.7\gev$. These curves are meant to be illustrative, based on
reasonable hypotheses: for $\btopi$ the results in~\cite{ukqcd:hlff}
were obtained for unphysically large $u$ and $d$ quark masses; for
$\b2rho$ we do not necessarily believe that the three form factors
involved are all simultaneously single pole types with the same pole
mass (the pole mass of $5.3\gev$ is an estimate of the average of the
$1^-$ and $1^+$ $b \bar u$ resonance masses). The points we wish to
make are:
\begin{enumerate}
\item For $\b2rho$ the events near $q^2=0$ have already been
recommended for study, but the expected number of events at large
$q^2$ is at least comparable.
\item In the region beyond the charm threshold, the rates for $\b2rho$
and $\btopi$ are at least comparable ($\btopi$ is not ten times more
common than $\b2rho$, for example), so both should be sizable
contributions to the inclusive $b\to u$ event rate and either or both
can be used for the determination of $\vub$.
\end{enumerate}

The lattice calculations of the form factors in the previous section
provide data points for $d\Gamma/dq^2$ over a sizable fraction of the
region in $q^2$ above the charm endpoint, as can be seen in
figure~\ref{fig:btorhovsbtopi}.  The uncertainties in determining
$A_2$ and $V$ in the $0\to0$ and $1\to0$ momentum channels do not
prejudice our measurement of the differential decay rate. The rate is
forced to vanish for kinematical reasons at $\qsqmax$ so the $0\to0$
channel form factor values of $A_2$ and $V$ are irrelevant.  Moreover,
the contribution to the differential decay rate of these two form
factors in the $1\to0$ channel is highly suppressed (the $1\to0$
channel has a $q^2$ near $\qsqmax$ and the $A_2$ and $V$ contributions
are suppressed relative to that of $A_1$ by a factor ${\bf k}^2/m_B^2$
as shown in equation~(\ref{eq:diffdecayrate})). Over the range of
$q^2$ for which we have measurements, the differential decay rate is
dominated by the form factor $A_1$. This is shown by the difference
between the points marked by squares and crosses in
figure~\ref{fig:btorhovsbtopi}.

In table~\ref{tab:diffdecaypoints} we give the numerical values we
have obtained for the differential decay rate at several $q^2$
values. We have parametrised the results of the table by fitting our
measured points for $d\Gamma/dq^2$ to the form
\begin{equation}\label{eq:param}
{10^{12}\over\vub^2}\frac{d\Gamma(\b2rho)}{dq^2} =
\frac{G_F^2}{192 \pi^3 m_B^3}
 q^2 [\lambda(q^2)]^{\frac{1}{2}} a^2 \big( 1 + b(q^2-\qsqmax) \big),
\end{equation}
obtained by retaining the phase space dependence and expanding $H^\pm$
and $H^0$ from equation~(\ref{eq:diffdecayrate}) around $\qsqmax$. The
results of the fit are
\begin{eqnarray}
a^2 &=& 21^{+3}_{-3} \gev^2, \nonumber\\
b   &=& (-8^{+4}_{-6}) 10^{-2} \gev^{-2},\label{eq:abvalues}
\end{eqnarray}
where the quoted errors are statistical. A 24\% systematic error
should be added as discussed below.

\begin{figure}
\hbox to\hsize{\hfill\epsfxsize=0.6\hsize
\epsffile[25 37 516 505]{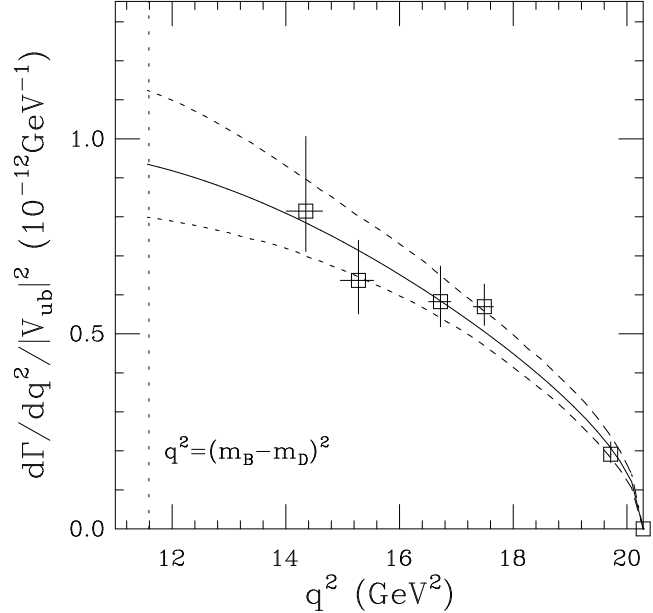}
\hfill}
\caption[]{Differential decay rate as a function of $q^2$ for the
semileptonic decay $\b2rho$. Squares are measured lattice data, solid
curve is fit from equation~\protect(\ref{eq:param}) with the parameters
given in equation~\protect(\ref{eq:abvalues}). Dashed lines are 68\%
confidence limits (statistical errors only). The vertical dotted line
marks the charm threshold.}
\label{fig:vub}
\end{figure}

With this parametrisation, we can partially integrate the differential
decay rate, giving the results shown in the last column of
table~\ref{tab:diffdecaypoints}. The table also includes the partially
integrated decay rate from the charm threshold to $\qsqmax$, although
higher order terms in the Taylor expansion in
equation~(\ref{eq:param}) could become sizable at the lower end of
this range, where we do not currently have measured points. The
results of table~\ref{tab:diffdecaypoints} and
equation~(\ref{eq:abvalues}) have been obtained by extrapolating the
form factors linearly in $1/M$. Quadratic extrapolations in $1/M$ give
results differing in the last significant figure, which is always well
within the statistical errors.  In figure~\ref{fig:vub} we show our
measured data together with the fit, including 68\% confidence level
bounds.
\begin{table}
\hbox to \hsize{\hfill\def\arraystretch{1.2}
\begin{tabular}{ccc}
\hline
\vrule height3ex width0pt depth0pt
$q^2$ & $d\Gamma/dq^2$ &
  $\int_{q^2}^{\qsqmax}dq^2\, d\Gamma/dq^2$ \\[5pt]
\hline
\tstrut
20.3 & 0.0 & 0.0 \\
19.7\er11 & 0.19\er32 & 0.08\er11\\
17.5\er22 & 0.57\er65 & 0.9\er11\\
16.7\er22 & 0.58\er96 & 1.3\er11\\
15.3\er33 & 0.6\er11 & 2.3\er22 \\
14.4\er33 & 0.8\er21 & 3.0\er32 \\[2pt]
\hline
\tstrut
{\small charm threshold} & & \\
11.6 & --- & 5.4\er75 \\[2pt]
\hline
\end{tabular}\hfill}
\caption[]{Differential and partially integrated decay rate for
$\b2rho$ measured on the lattice for various $q^2$ values. $q^2$ is
given in units of $\gev^2$ and $\Gamma$ in units of $\vub^2 10^{-12}
\gev$. For the $0\to0$ channel, $\w$ is $1$ exactly, so that $q^2$ is
fixed up to (tiny) experimental errors in the physical meson masses.
The integration has been performed using the parametrisation of
equation~\protect(\ref{eq:param}). Quoted errors are purely
statistical --- a further 24\% systematic error should be added as
discussed in the text.}
\label{tab:diffdecaypoints}
\end{table}

Comparison of any of the values in table~\ref{tab:diffdecaypoints}
with experimental measurements will allow an extraction of $\vub$ with
less than 10\% statistical and 12\% systematic theoretical
uncertainties. Another way to determine $\vub$ uses the
parametrisation of equation~(\ref{eq:param}) and our results of
equation~(\ref{eq:abvalues}). It parallels the method used to extract
$|V_{cb}|$ from semileptonic $\bar B \to D^*$ decays (see
references~\cite{neubert:ichep94}--\cite{argus:btodstar}).  The factor
$a^2 \big(1 + b(q^2-\qsqmax)\big)$ in equation~(\ref{eq:param})
parametrises long distance hadronic dynamics (it is the analogue of
the function $\hat\xi^2(\w)$ in~\cite{neubert:ichep94} for the $\bar B
\to D^*$ case). The idea is to use experimental measurements of
$d\Gamma/dq^2$ extrapolated to $\qsqmax$ to extract a value for
$\vub^2 a^2$ and compare with our theoretical determination of
$a^2$. Here $a$ plays the role of $\hat\xi(1)$ in the $\bar B \to D^*$
extraction of $|V_{cb}|$.

The lattice simulation provides a systematic determination of the
overall normalisation of hadronic dynamical effects, parametrised by
$a$. For the extraction of $|V_{cb}|$ from $\bar B \to D^*$ decays HQS
provides such a normalisation (the Isgur-Wise function is $1$ at zero
recoil, up to $1/M^2$ and short distance perturbative
corrections). For the heavy-to-light $\b2rho$ semileptonic decay, such
a normalisation is not provided by HQS.

In the fit to equation~(\ref{eq:param}) the $0\to0$ channel is not
used because $d\Gamma/dq^2|_{\qsqmax}$ is zero and our fitting
function automatically incorporates this feature. Our value for
$A_1(\qsqmax)$ from the $0\to0$ channel therefore gives an independent
measurement of $a^2$ because $V$ and $A_2$ do not contribute at this
kinematic point. We find, using the value from
table~\ref{tab:formfactorvalues}, and quoting statistical errors only,
\begin{equation}\label{eq:a2from00}
a^2 = 3(m_B + m_\rho)^2 A_1^2(\qsqmax) = 23^{+2}_{-2} \gev^2,
\end{equation}
in excellent agreement with $a^2$ determined by the fit to the other
channels.  This suggests that discretisation and other systematic
errors have not conspired to change drastically the shape of the
overall $q^2$ dependence. Our final result will use the value of
$a^2$, given in equation~(\ref{eq:abvalues}), determined from the fit
to these other channels, because this does not rely on a single
measurement.

To estimate systematic uncertainties, we have considered four possible
sources of errors: quenching, determination of the lattice spacing in
physical units, light spectator mass independence of the form factors
and discretisation errors.

The exact effect of ignoring internal quark loops is an unknown
systematic error. However, quenched lattice calculations of form
factors for $D\to K,K^*$ semileptonic decays (see~\cite{ukqcd:dtok}
and references therein) give results in agreement with world average
experimental values, while quenched calculations of $\bar B \to
D,D^*$~\cite{ukqcd:btodstar} decays have allowed extractions of
$|V_{cb}|$ in agreement with other determinations of that quantity.
This gives us some confidence that errors due to quenching are most
likely within the statistical errors for processes involving a heavy
quark.

The error arising from uncertainties in the value of the lattice
spacing in physical units should be minimised in our calculation,
because we have evaluated only dimensionless quantities: form factors
and values of the velocity transfer, $\w = v\cdot v'$. Physical meson
masses have been used as input where dimensional results are required.
The only dependence on the lattice scale enters in expressing the
heavy-light pseudoscalar masses in physical units for the $1/M$
extrapolations of the form factors and in the short distance
logarithmic corrections of equation~(\ref{eq:theta}). The systematic
error induced is much smaller than the other errors we will consider
below.

We believe our results show good evidence for light spectator mass
independence of the form factors. We made this independence an
assumption in order to use the results for the heaviest light
spectator quark only: these results have smaller statistical error and
are available for all our heavy quarks. In the previous section we
argued that we could not rule out a 5\% error in $A_1$ (10\% in
channels involving $\sqrt2$ momenta for the light meson) due to this
assumption. The corresponding errors in $A_2$ and $V$ were 20\% and
10\% respectively.  In the region near $\qsqmax$, $d\Gamma/dq^2$ is
dominated by $A_1$ and the contributions of $V$ and $A_2$ are
kinematically suppressed, being 5\% or less for $q^2 > 16\gev^2$. The
$0\to\sqrt2$ and $1\to\sqrt2$ channels explore lower $q^2$ values
where the $V$ and $A_2$ contributions are less suppressed, of order
15--20\% (see figure~\ref{fig:btorhovsbtopi}). If the spectator quark
mass independence assumption were violated by as much as 20\% for $V$
and $A_2$, the error in the differential decay rate would only be of
the order of 2\% for the non-$\sqrt2$ channels, and 6--8\% for the
$\sqrt2$ channels noted above, smaller in both cases than the likely
10\% error induced by the same assumption for $A_1$ (the differential
decay rate is proportional to the square of the form
factors)\footnote{We have studied the effect on the results of the fit
of equation~(\protect\ref{eq:abvalues}) of having a 10\% error in the
$\sqrt2$ light meson momentum channels. This changes the parameter~$a$
by 5\%.}. Systematic errors due to this approximation are within our
statistical errors. A conservative estimate of the systematic error in
the differential decay rate arising from this source can be obtained
by adding in quadrature the 10\% error due to $A_1$ and the worst-case
8\% error from $V$ and $A_2$, to get a final error of 13\%. We will
see below that discretisation errors are expected to be almost twice
as large as this.

We discussed in section~\ref{sec:latticedetails} that we could not
dismiss having 10\% errors in the form factors arising from
discretisation errors at the value of $\beta$ used. In consequence, we
will admit a 20\% error in the decay rate which translates into a 10\%
error in $\vub$. These errors are the largest among the systematic
effects considered, and they almost entirely determine our overall
systematic error. Combining errors from discretisation and our
assumption of light quark spectator mass independence of the form
factors in quadrature, we finally obtain an estimated 24\% systematic
error in the decay rate, which should be added to the results of
table~\ref{tab:diffdecaypoints} and
equations~(\ref{eq:param})--(\ref{eq:a2from00}).

\section{Conclusions}

We have presented a method for extracting the CKM matrix element
$\vub$ from experimental measurements of the exclusive decay $\b2rho$.
We have shown how lattice simulations provide a model-independent
framework in which nonperturbative QCD corrections can be
systematically incorporated in the evaluation of the differential
decay rate $d\Gamma(\b2rho)/dq^2$ in a region of $q^2$ values from the
charm threshold up to the zero recoil point, $\qsqmax$. We have
measured points spanning roughly two-thirds of this region of
$q^2$. One of the main features of the approach presented here is that
it does not rely upon large, difficult and model-dependent
extrapolations in $q^2$ from $\qsqmax$ down to values close to $q^2 =
0$, where the semileptonic decays are dominated by charmed final
states, as in other previously suggested methods.

Discounting experimental errors, the results of
table~\ref{tab:diffdecaypoints} will allow the determination of the
CKM matrix element $\vub$ with a theoretical uncertainty of less than
10\% statistical and 12\% systematic. Alternatively, determinations of
$\vub$ (with similar theoretical ambiguities) can be obtained from the
overall normalisation of the hadronic effects parametrised by $a$,
found to be
\begin{equation}
a = 4.6 {}^{+0.4}_{-0.3} {\rm\ (stat)} \pm 0.6 {\rm\ (syst)} \gev.
\end{equation}
Both statistical and systematic theoretical errors will be reduced in
forthcoming lattice simulations with smaller lattice spacing (to
reduce discretisation errors, currently the principal systematic
effect) and higher statistics.

Currently, only an upper bound for the total decay rate
$\Gamma(\b2rho)$ is known experimentally, but new results are
forthcoming. It is important that the differential decay rate near
$\qsqmax$ be measured, since, as we have shown, such a measurement can
provide a clean determination of $\vub$ in the same way that similar
measurements for the $\bar B\to D^*$ semileptonic decay have
successfully been used to extract an accurate value for $|V_{cb}|$.

In this paper we have also determined the $q^2$ dependence of the form
factor $A_1$ (which dominates the decay rate close to the zero recoil
point) for the $\b2rho$ decay in the region from $q^2=0$ to
$\qsqmax$. The form factors $V$ and $A_2$, for the same semileptonic
decay, have been studied in a region of $q^2$ around $\qsqmax$.

Finally, relations between radiative and semileptonic $\bar B$ decays
into light vector mesons, predicted by HQS, have been examined and
found to hold within errors for the ratio $A_1/2iT_2$ and within 20\%
for the ratio $V/2T_1$ at the $B$ scale.

Using extra momentum channels where the heavy-light meson is not at
rest, but $\w$ is nevertheless nearly constant as the heavy quark mass
changes~\cite{ukqcd:hlff}, has been essential for studying $q^2$, or
equivalently $\w$, dependences of $A_1$, $A_1/2iT_2$, $V/2T_1$ and
$d\Gamma/dq^2$. Knowledge of the $q^2$ dependence of $d\Gamma/dq^2$
has allowed us to determine the parameter $a$ from a set of five
measurements as well as from the zero recoil channel, giving us
further confidence in our procedure for extracting $\vub$.

\subsection*{Acknowledgements}

We thank Chris Sachrajda and Hartmut Wittig for useful discussions.
This research was supported by the UK Science and Engineering Research
Council under grants GR/G 32779 and GR/H 49191, by the Particle
Physics and Astronomy Research Council under grant GR/J 21347, by the
University of Edinburgh and by Meiko Limited.
We are grateful to Edinburgh University Computing Service and, in
particular, to Mike Brown, for maintaining service on the Meiko i860
Computing Surface.
DGR acknowledges the Particle Physics and Astronomy Research Council
for support through an Advanced Fellowship.
%
We acknowledge the Particle Physics and Astronomy Research Council for
travel support under grant GR/J 98202.

\end{document}